\documentclass[11pt]{article}

\usepackage[top=1.0in,bottom=1.0in,left=1.0in,right=1.0in]{geometry}
\usepackage{bigints}
\usepackage{amsmath,amssymb,amsthm}
\usepackage{graphicx}
\usepackage{float}
\usepackage{stmaryrd} 
\usepackage{relsize} 
\usepackage{algorithm}
\usepackage{algpseudocode}
\usepackage{color}
\usepackage[dvipsnames]{xcolor}
\usepackage{setspace}
\usepackage{enumitem}
\usepackage[normalem]{ulem} 
\usepackage[font=footnotesize]{caption}
\usepackage{hyperref}
\hypersetup{
    colorlinks=true,
    linkcolor=magenta,
    urlcolor=blue,
    citecolor=blue
}

\newcommand{\nx}{N_x}
\newcommand{\nt}{N_t}

\newcommand{\htt}{h_t}

\newcommand{\Lag}{\mathcal{L}}
\newcommand{\Act}{\mathcal{S}}
\newcommand{\Ham}{\mathcal{H}}
\newcommand{\Slag}{\mathcal{S}_{\ell}}
\newcommand{\Tlag}{\mathcal{T}_{\ell}}
\newcommand{\Sham}{\mathcal{S}_{h}}
\newcommand{\Tham}{\mathcal{T}_{h}}
\newcommand{\Shamq}{\mathcal{S}_{h \mu}}
\newcommand{\Thamq}{\mathcal{T}_{h \lambda}}
\newcommand{\Shamp}{\mathcal{S}_{h \nu}}
\newcommand{\Thamp}{\mathcal{T}_{h \omega}}

\newcommand{\Imega}{\mathcal{I}}

\newcommand{\midsum}{\mathlarger{\mathlarger{\sum}}}

\newcommand{\myint}{\bigintssss}

\newcommand{\tr}{^{T}} 

\newcommand{\Mb}{\bar{M}}
\newcommand{\Mbb}{\bar{\bar{M}}}

\newcommand{\ps}{p^{\ast}}
\newcommand{\qs}{q^{\ast}}
\newcommand{\qsp}{\dot{q}^{\ast}}

\newcommand{\qp}{\dot{q}}
\newcommand{\qpp}{\ddot{q}}
\newcommand{\ppoint}{\dot{p}}
\newcommand{\lambdap}{\dot{\lambda}}
\newcommand{\lambdapp}{\ddot{\lambda}}
\newcommand{\lambdas}{\lambda^{\ast}}
\newcommand{\mus}{\mu^{\ast}}
\newcommand{\nus}{\nu^{\ast}}
\newcommand{\omegas}{\omega^{\ast}}

\newcommand*{\virt}[2]{%
  \vbox{%
    \kern0.25ex
    \hbox{%
      \kern-0.1em
      \ifmmode #1^{\ast} \else\ensuremath{#1^{\ast}}\fi
      \kern-0.1em
    }
  }^{\hspace{0.1em}#2}
}

\newcommand*{\virtbar}[2]{%
  \vbox{%
    \hrule height 0.5pt
    \kern0.25ex
    \hbox{%
      \kern-0.1em
      \ifmmode #1^{\ast} \else\ensuremath{#1^{\ast}}\fi
      \kern-0.1em
    }
  }^{\hspace{0.1em}#2}
}

\newcommand{\llambda}{\boldsymbol{\lambda}}

\newcommand{\oomega}{\boldsymbol{\omega}}

\newcommand{\rr}{\boldsymbol{r}}

\newcommand{\Q}{Q}
\newcommand{\Qp}{\dot{\Q}}
\newcommand{\Qpp}{\ddot{\Q}}
\newcommand{\Qs}{Q^{\ast}}
\newcommand{\PP}{P}
\newcommand{\Pp}{\dot{P}}
\newcommand{\Ps}{P^{\ast}}
\newcommand{\F}{F}
\newcommand{\W}{W}

\newcommand\clem[1]{{\color{black}{#1}}}

\begin{document}

\title{PGD Reduced-Order Modeling for Structural Dynamics Applications}

\author{Cl\'ement Vella\textsuperscript{a}, Serge Prudhomme\textsuperscript{b,}\thanks{Corresponding author} \vspace{.1in} \\
\footnotesize \textsuperscript{a} LaMCube (UMR 9013), Universit\'e de Lille,\\
\footnotesize Avenue Paul Langevin, 59655 Villeneuve d'Ascq, France \\
\footnotesize \textsuperscript{b} Department of Mathematics and Industrial Engineering, Polytechnique Montr\'eal,\\
\footnotesize C.P. 6079, succ.\ Centre-ville, Montr\'eal, Qu\'ebec H3C~3A7, Canada
}

\makeatletter
\def\blfootnote{\gdef\@thefnmark{}\@footnotetext}
\makeatother

\blfootnote{\hspace{-.09in}\textit{E-mail addresses:} \url{clement.vella@univ-lille.fr} (C. Vella), \url{serge.prudhomme@polymtl.ca} (S. Prudhomme) \\ \indent \textit{URL:} \url{https://www.polymtl.ca/expertises/en/prudhomme-serge} (S. Prudhomme)}

\date{}

\maketitle

\noindent\rule{\textwidth}{0.5pt}
\vspace{-1cm}
\section*{Abstract}
We propose in this paper a Proper Generalized Decomposition (PGD) approach for the solution of problems in linear elastodynamics. The novelty of the work lies in the development of weak formulations of the PGD problems based on the Lagrangian and Hamiltonian Mechanics, the main objective being to devise numerical methods that are \clem{numerically} stable and energy conservative. We show that the methodology allows one to consider the Galerkin-based version of the PGD and numerically demonstrate that the \clem{PGD solver based on the} Hamiltonian formulation offers better stability and energy conservation properties than the Lagrangian formulation. 
The performance of the two formulations is illustrated and compared on several numerical examples describing the dynamical behavior of a one-dimensional bar.

\vspace{.1in}
\noindent\textit{Keywords:} Proper Generalized Decomposition, Wave Equation, Lagrangian and Hamiltonian Mechanics, Law of Conservation of Energy

\noindent\rule{\textwidth}{0.5pt}

\section{Introduction}
\label{sect:introduction}

Applications in structural dynamics often require fast methods to efficiently estimate solutions for real-time simulations such as in digital twins, for uncertainty quantification analyses or multi-query optimization. 
In order to do so, we are interested here in constructing \textit{a priori} reduced-basis representations of the solutions to a second-order hyperbolic system that describes the linear elastodynamics behavior of a bar, basically the propagation of a wave in a one-dimensional medium. 
There exist several approaches for model-order reduction, for instance the Proper Orthogonal Decomposition (POD)~\cite{Amsa08,Gunz07}, the Proper Generalized Decomposition (PGD)~\cite{chinesta,nouy}, the Reduced-Basis (RB)~\cite{Boyaval2010,Rozza08} methods, among others. The reduced-order method that we consider here is the Proper Generalized Decomposition (PGD) approach, which provides a straightforward method to build a reduced basis on-the-fly, without any \textit{a priori} knowledge of the solution to the problem at hand. The PGD method is akin to the method of separation of variables, in the sense that one assumes that the solutions to differential equations are separable with respect to the independent variables and/or the model parameters. We will assume in this work that the solutions to the wave equation are space-time separable, in other words, that they can be approximated by a sum of products of functions in space by functions in time. The space and time functions, which will be referred to as the modes, are the unknowns of the PGD formulation.

Several efforts have been deployed in recent years towards building separated approximations of solutions governed by second-order hyperbolic equations. In general, methods differ from each other depending on the form of the solutions that are chosen to describe such systems, on the choice of the PGD formulation  used to build the reduced model, or on the numerical scheme used for the discretization in time.
Low-rank separable representations in space and time for elastodynamics were proposed and extensively studied in~\cite{boucinha2013,boucinha2014} even though it was known that constructing these representations in an efficient manner could sometimes be challenging~\cite{ladeveze,ammar2010PGD}. The authors considered either the displacement alone or both the displacement and velocity as the unknown fields. The first approach led to a PGD formulation in which only a separated representation for the displacement was considered while the second approach introduced two distinct representations, one for the displacement and one for the velocity, and was called the Multi-Field PGD (MF-PGD). They also  investigated the Galerkin-based and the minimal residual versions of the PGD. Numerical experiments showed that their Galerkin-based version would diverge while the minimal residual-based version would consistently converge for all test cases and time integration schemes that were implemented, namely the Newmark, Time Continuous Galerkin (TG), and Time Discontinuous Galerkin (TDG) methods. In fact, the minimal residual-based version of the PGD was proven in~\cite{ammar2010TC} to converge in the case of greedy rank-one update algorithms.
In~\cite{bamer}, the authors developed a space-time representation using a PGD approach based on a non-incremental Newmark integration scheme. They considered the discretized wave equation, both in space and time, within the Newmark equation to obtain an algebraic system. Then, the fully discrete space-time separated representation of the displacement (rank two tensor) was introduced into the algebraic system, which was alternatively solved for the spatial and the temporal modes in a fixed point procedure.
\clem{Different approaches have also been proposed in order to circumvent the difficulty one encounters when using a separated representation in space and time. For instance, an approach that assumes a good space separation, was presented in~\cite{goutaudier}. It consists in estimating, in an adaptive manner, the number of spatial modes at each time step. Alternatively, space-frequency separated representations were developed for a fairly wide range of applications: 2D acoustics~\cite{barbarulo, debrabander2022}, non-linear soil mechanics~\cite{germoso2016}, linear and non-linear structural dynamics~\cite{malik2018-1, quaranta2019} and transient electronics~\cite{malik2018-2}. The space-frequency separation was shown to be particularly efficient if additional parameters (material, geometric, etc.) were accounted for. PGD has also been successfully used in the context of structural dynamics with the objective of building parametric solutions for industrial problems in a non-intrusive manner~\cite{cavaliere2021, cavaliere2022}.}

In this paper, we develop novel space-time separated representations for the wave equation with the objective of devising numerical methods that are stable and energy conservative. Particular attention is paid to the derivation of weak formulations, at the continuous level, following the application of the Hamilton's Weak Principle~\cite{nasahwp}. The first formulation is based on the Lagrangian description of Mechanics~\cite{lagrange}, where only the displacement field is considered as an unknown of the problem. The second one is based on the Hamiltonian theory~\cite{hamilton1834, hamilton1835}, in which both the displacement field (generalized coordinates) and the conjugate field (generalized momenta) are treated as unknown fields. We thus derive two weak formulations that allow us to implement the Galerkin-based version of the PGD, which will be referred to as L-PGD and H-PGD, respectively. In particular, the Hamiltonian approach naturally leads to a mixed weak formulation that allows one to introduce two separated representations, one for the displacement field and the other for the conjugate field, in a manner similar to the MF-PGD~\cite{boucinha2013}, but using the Galerkin-based PGD.

Regarding the discretization in time, several integration schemes have been adapted to linear elastodynamics~\cite{hughesfem, hulbert, rixen}. Only stable, energy conservative discretization schemes are considered here, namely the Crank-Nicolson method~\cite{estep, nandy, iavernaro} (also known as the implicit trapezoidal rule), and the Newmark method~\cite{newmark} with $\gamma=1/2$ and $\beta=1/4$. Moreover, we apply two post-processing procedures that aim at improving the convergence of the Galerkin-based version of the PGD~\cite{nouy, alameddin}, namely 1) the orthogonalization of the spatial modes via a modified Gram-Schmidt algorithm, and 2) the update procedure of the temporal modes.
These procedures are applied to both the L-PGD and H-PGD. We will show in the numerical examples that the H-PGD solver has a better numerical stability and produces solutions with better energy conservation than the L-PGD, and this for all implemented test cases. One reason is that the orthonogonalization and update procedures for the H-PGD truly work in synergy. Moreover, we propose an adaptive fixed-point algorithm for the H-PGD that independently controls, and thus accelerates, the convergence of the fields. Finally, we will show through a numerical example that the methodology can be extended in a straightforward manner to the case of the wave equation involving a linear damping term.


The paper is organized as follows: in Section~\ref{sect:modelproblem}, we describe the model problem and provide an analytical solution by the method of separation of variables for a specific set of initial and boundary conditions. In Section~\ref{sect:weakformulations}, we present the weak formulations of the problem based on the Lagrangian and Hamiltonian formalism and derive discrete counterparts using the Finite Element method in space and numerical integration schemes in time. The L-PGD and H-PGD approaches are described in Section~\ref{sect:PGD} along with the orthogonalization and updating procedures as well as the fixed-point algorithms.
Numerical experiments are presented in Section~\ref{sect:numerical_results} to illustrate the performance of the proposed approaches. We finally provide some concluding remarks in Section~\ref{sect:conclusions}.

\section{Model problem}
\label{sect:modelproblem}

\subsection{Strong formulation}

The model problem we shall consider consists of a 1D bar in traction or compression under the assumption of infinitesimal deformation. The bar has density~$\rho$, Young's modulus~$E$, length~$\ell$, and cross-sectional area~$A$. We will assume that $E$ and $A$ are constant but that $\rho$ could possibly vary in space. Let $\Omega = (0, \ell)$ be the open interval in $\mathbb R$ occupied by the bar and let $\Imega = (0, T)$ denote the time interval. The displacement $u=u(x,t)$ is governed by the 1D wave equation:
\begin{equation}
\label{eq:wave_equation}
\rho A \frac{\partial^{2} u}{\partial t^{2}} - EA \frac{\partial^{2} u}{\partial x^{2}} = f, \qquad \forall (x,t) \in \Omega \times \Imega,
\end{equation}
and subjected to the initial conditions:
\begin{alignat}{2}
\label{eq:ICu}
u(x,0) &= u_{0}(x), &\qquad \forall x \in \Omega, \\
\label{eq:ICv}
\frac{\partial u}{\partial t}(x,0) &= v_{0}(x), &\qquad \forall x \in \Omega,
\end{alignat}
as well as to the boundary conditions:
\begin{alignat}{2}
\label{eq:DirichletBC}
u(0,t) &= 0, &\qquad \forall t \in \Imega, \\
\label{eq:NeumannBC}
EA \frac{\partial u}{\partial x} (\ell,t) &= g(t), &\qquad \forall t \in \Imega,
\end{alignat}
where the functions $f=f(x,t)$, $u_0=u_0(x)$, $v_0=v_0(x)$, and $g=g(t)$ are supposed sufficiently regular to yield a well-posed problem. In the following, we will denote the time derivatives by $\dot{u} = \partial u /\partial t$ and $\ddot{u} = \partial^2 u/\partial t^2$ and the space derivatives by $u'=\partial u/\partial x$ and $u''=\partial^2 u/\partial x^2$. Moreover, we introduce the wave speed as $c=\sqrt{E/\rho}$.

\subsection{Analytical solution}

In the case that the speed $c$ is chosen constant, it is well known that the general solution to the homogeneous wave equation~\eqref{eq:wave_equation} in an infinite domain, i.e.\ with $f(x,t)=0$ and $\Omega=\mathbb R$, can be recast, using the d'Alembert formula, as $u(x, t) = \varphi(x+ct) + \phi(x-ct)$, where $\varphi$ and $\phi$ are identified from the initial data $u_0$ and $v_0$. The solution is therefore interpreted as two waves with constant velocity $c$ moving in opposite directions along the $x$-axis. In the particular case that~$v_0(x) = 0$, the solution is given by $u(x,t) = u_0(x+ct) + u_0(x-ct)$. It follows that the solution may not always be represented in a separated form with respect to both space and time depending on the choice of~$u_0$.  

We nevertheless provide the analytical solution in the case of a simple problem, that is, taking $f(x,t)=0$, $v_0(x)=0$, $g(t)=0$, \clem{and $c$ constant}. Moreover, the initial condition on the displacement is chosen as $u_0(x) = F x / (EA)$, which corresponds to the equilibrium state of the bar when subjected to a force $F$ at $x=\ell$. The displacement $u$ satisfies the following system of equations:
\begin{align}
\label{eq:hom_wave_equation}
\ddot{u} - c^{2} u'' = 0, 
&\qquad \forall (x,t) \in\Omega\times\Imega,\\
u(x,0) = u_0(x), 
&\qquad \forall x \in \Omega, \nonumber\\
\dot{u}(x,0) = 0,
&\qquad \forall x \in \Omega, \nonumber\\
u(0,t) = 0,
&\qquad \forall t \in \Imega, \nonumber\\
EA u'(\ell,t) = 0, 
&\qquad \forall t \in \Imega. \nonumber
\end{align}
Using the method of separation of variables, we search for solutions in the separated form:
\begin{equation*}
u(x,t) = \chi(x) \psi(t).
\end{equation*}
Substituting the above expression for $u$ in~\eqref{eq:hom_wave_equation} yields:
\begin{equation*}
\frac{\chi''(x)}{\chi(x)} = 
\frac{1}{c^2}\frac{\ddot{\psi}(t)}{\psi(t)},
\qquad \forall (x,t) \in \Omega \times \Imega.
\end{equation*}
It follows that one has to find constants $\lambda \in \mathbb{R}$ such that the function $\chi(x)$ satisfies the eigenvalue problem:
\begin{equation}
\label{eq:sous_pb}
\begin{aligned}
\chi''(x) + \lambda \chi(x) = 0, 
&\qquad \forall x \in \Omega,\\
\chi(0) = 0,\\
\chi'(\ell) = 0,
\end{aligned}
\end{equation}
and such that the function $\psi(t)$ satisfies the ordinary differential equation:
\begin{equation}
\ddot{\psi}(t) + c^2 \lambda \psi(t) = 0, 
\qquad \forall t \in \Imega. \label{eq:psi}
\end{equation}
The solutions to the eigenvalue problem~\eqref{eq:sous_pb} consist of the eigenvalues~$\lambda_k$ and associated eigenfunctions~$\chi_k$:
\begin{equation*}
\lambda_k = \bigg[ \frac{(2k-1)\pi}{2\ell} \bigg]^2,
\qquad 
\chi_k(x) = \sin \left( \sqrt{\lambda_k} x \right),
\qquad \forall k=1,2,\ldots
\end{equation*}
while the solution to~\eqref{eq:psi} for each eigenvalue $\lambda_k$ is given as:
\begin{equation*}
\psi_k(t) = \alpha_k \cos \left( c\sqrt{\lambda_k}t \right) + \beta_k \sin  \left(c\sqrt{\lambda_k}t \right).
\end{equation*}
The general solution to the problem thus reads: 
\begin{equation*}
u(x,t) 
= \sum_{k=1}^{+\infty} \chi_k(x) \psi_k(t)
= \sum_{k=1}^{+\infty} \sin \frac{(2k-1)\pi x}{2\ell} \left[ \alpha_{k} \cos\frac{(2k-1)\pi ct}{2\ell} + \beta_{k} \sin\frac{(2k-1)\pi ct}{2\ell} \right].
\end{equation*}
Using the initial condition on the velocity, i.e.\ $\dot{u}(x,0) = v_0(x) = 0$, implies that $\beta_{k}=0$ for all $k=1,2,\ldots$ Moreover, the coefficients $\alpha_k$ correspond to the coefficients of the sine Fourier series associated with the initial displacement $u_0(x)=Fx/(EA)$.
It follows that the displacement field reads:
\begin{equation}
\label{eq:analytic_equation}
u(x, t) = \frac{8F\ell}{\pi^{2}EA} 
\sum_{k=1}^{+\infty} \frac{(-1)^{k+1}}{(2k-1)^{2}} 
\sin\frac{(2k-1)\pi x}{2\ell} \cos\frac{(2k-1)\pi ct}{2\ell}, \qquad \forall (x,t) \in \bar{\Omega} \times \bar{\Imega}. 
\end{equation}
We observe in this case that the solution can be represented in a separated form and that the coefficients of the series decrease with a quadratic rate. We shall use this analytical solution to assess the accuracy of our calculations in some of the numerical examples.

\section{Weak formulations of the problem}
\label{sect:weakformulations}

The construction of weak formulations of the problem is not unique. We present below two formulations based on the Lagrangian and the Hamiltonian approaches. We first recall the Hamiltonian's principle that will be used in the derivation.

\subsection{Hamilton's Weak Principle}

Let $q$ denote the generalized coordinates of the system, corresponding here to the displacement field $u$, and let $t_{A}$ and $t_{B}$ be two specified times. \clem{We note that the principle was originally stated assuming that the initial and final states of the system under study were known}, $q(x, t_{A}) = q_{A}(x)$ and $q(x, t_{B}) = q_{B}(x)$. Given a Lagrangian functional~$\Lag$ of the system, the action functional of $q$, denoted by $\Act[q]$, is defined as~\cite{hamilton1834, hamilton1835}: 
\begin{equation}
\label{eq:action}
\Act[q] = \myint_{t_{A}}^{t_{B}}{ \Lag(q(t), \qp(t), t) \, dt}.
\end{equation}
The Hamilton's Weak Principle states that the evolution of $q$ followed by the physical system between the states $q_{A}$ and $q_{B}$ is a stationary point of the action functional:
\begin{equation}
\Act'[q](\qs) = 0, \qquad \forall \qs \in V_0,
\end{equation}
where $V_0$ is the space of perturbations $\qs$ that vanish at $t_{A}$ and $t_{B}$. The precise definition of $V$ depends on the choice of the Lagrangian. Here, $\Act'[q](\qs)$ is the G\^ateaux derivative of $\Act[q]$ defined at $q$ with respect to the perturbation $\qs$, i.e.
\begin{equation*}
\Act'[q](\qs) = \lim_{\theta\rightarrow 0} \frac{1}{\theta} \big(\Act[q+\theta\qs] - \Act[q]\big).
\end{equation*}
In the particular case where the states $q$ at times $t_{A}$ and $t_{B}$ are unknown, the principle of least action can be recast as \cite{baruch}:
\begin{equation}
\label{eq:general_hwp}
\Act'[q](\qs) = \left[ \frac{\partial \Lag}{\partial \qp} \qs \right]_{t_{A}}^{t_{B}}, \qquad \forall \qs \in V. 
\end{equation}
We note here that the perturbations $\qs$ in $V$ do not necessarily vanish at $t_{A}$ or $t_{B}$. \clem{Later in the manuscript, all our test cases consider the initial displacement $q_{A}$ to be known while $q_{B}$ remains unknown.}

\subsection{The Lagrangian formalism}

The evolution of the generalized coordinates function $q$, describing the displacement as a function of $x$ and $t$, defines a so-called trajectory of a system in the configuration space. The trajectory of a physical system is thus entirely determined by the knowledge of $q$. The objective of Joseph-Louis Lagrange in his seminal treatise \textit{Méchanique Analitique} first published in 1788~\cite{lagrange} was to lay down, once and for all, the foundations of analytical mechanics. In fact, he introduced as early as 1756 the action functional $\Act$ as defined in~\eqref{eq:action}. 

\subsubsection{Continuous formulation}

For our problem~\eqref{eq:wave_equation}-\eqref{eq:NeumannBC}, the Lagrangian functional~$\Lag$ reads:
\begin{equation}
\label{eq:lagrangian}
\Lag(q, \qp, t) 
= \underbrace{ \frac{1}{2}\myint_{\Omega}{\rho A \qp^{2}\, dx} }_{\text{Kinetic Energy}} 
- \underbrace{ \frac{1}{2}\myint_{\Omega}{EA \left( \frac{\partial q}{\partial x} \right)^{2} dx} }_{\text{Potential Energy}} 
+ \underbrace{ \myint_{\Omega}{fq\ dx} + g(t) q(\ell,t) }_{\text{External Energy}},
\end{equation}
where the field $q$ satisfies the initial conditions~\eqref{eq:ICu}-\eqref{eq:ICv}. In other words, the space of trial fields $q$ is:
\begin{equation}
\begin{aligned}
&\mathcal U_L = \{ q \in L^2(\Imega, H^1(\Omega)) \cap H^1(\Imega, L^2(\Omega));
\\
& \hspace{1.0in}
q(0,t)= 0,\ \forall t \in \Imega;\ q(x,0) = u_0(x),\ \qp(x,0) = v_0(x),\ \forall x\in \Omega \}.
\end{aligned}
\end{equation}
Requiring that the trajectory $q$ be a stationary point of the action functional~$\Act$ leads to the so-called Euler-Lagrange equations. Using the Lagrangian~\eqref{eq:lagrangian}, the G\^ateaux derivative of~$\Act$ is given by:
\begin{equation*}
\label{eq:derivativeS}
\begin{aligned}
\Act'[q](\qs) 
&= \lim_{\theta\rightarrow 0} \frac{1}{\theta} 
\bigg[ \int_{\Imega} \Lag(q +\theta \qs, \qp +\theta \qsp, t)\, dt - \int_{\Imega} \Lag(q, \qp, t)\, dt \bigg] \\
&= \lim_{\theta\rightarrow 0} \frac{1}{\theta} 
\int_{\Imega} \Big[ \Lag(q +\theta \qs, \qp +\theta \qsp, t) - \Lag(q, \qp, t) \Big] dt \\
&= \myint_{\Imega}{ \myint_{\Omega}{ \rho A \qsp \qp - EA \frac{\partial \qs}{\partial x} \frac{\partial q}{\partial x} + \qs f \, dx } dt} + \myint_{\Imega}{ \qs(\ell, t) g(t) \, dt},
\end{aligned}
\end{equation*}
where the space of perturbations $\qs$ is given by:
\begin{equation}
\label{eq:VLagrange}
\begin{aligned}
&\mathcal V_L = \{ \qs \in L^2(\Imega,H^1(\Omega)) \cap H^1(\Imega,L^2(\Omega));
\\
&\hspace{1.2in}
\qs(0,t)= 0,\ \forall t \in \Imega;\ \qs(x,0) = 0,\ \qsp(x,0) = 0,\ \forall x\in \Omega \}.
\end{aligned}
\end{equation}
Then, using~\eqref{eq:general_hwp}, we obtain the equation:
\begin{equation*}
\myint_{\Imega}{\myint_{\Omega}{ \rho A \qsp \qp - EA \frac{\partial \qs}{\partial x} \frac{\partial q}{\partial x} + \qs f \, dx } dt} + \myint_{\Imega}{ \qs(\ell, t) g(t) \ dt}
= \bigg[ \myint_{\Omega}{ \rho A \qs \qp\, dx} \bigg]_{0}^{T}, \quad \forall \qs \in \mathcal V_L.
\end{equation*}
It follows that a weak formulation of the problem reads:
\begin{equation}
\label{eq:weak-lagrange}
\begin{aligned}
&\text{Find $q \in \mathcal U_L$ such that}\\
&\qquad\quad 
\myint_{\Imega}{\myint_{\Omega}{ \rho A \qsp \qp - EA \frac{\partial \qs}{\partial x} \frac{\partial q}{\partial x}\, dx } dt} - \myint_{\Omega}{ \rho A \qs(x,T) \qp(x,T)\, dx} \\
&\hspace{2.5in} 
= - \myint_{\Imega}{ \myint_{\Omega}{ \qs f \, dx } dt} 
- \myint_{\Imega}{ \qs(\ell, t) g(t) \, dt}, 
\quad \forall \qs \in \mathcal V_L.
\end{aligned}
\end{equation}
We note that the above formulation is equivalent to the strong form~\eqref{eq:wave_equation}-\eqref{eq:NeumannBC} of the problem for sufficiently smooth data. Indeed, integration by parts with respect to the time variable yields:
\begin{equation}
\label{eq:pre_lag_hwp}
\myint_{\Imega}{ \myint_{\Omega}{ \rho A \qs \qpp + EA \frac{\partial \qs}{\partial x} \frac{\partial q}{\partial x} \, dx } dt} = \myint_{\Imega}{ \myint_{\Omega}{ \qs f \, dx } dt} + \myint_{\Imega}{ \qs(\ell, t) g(t) \, dt}, \quad \forall \qs \in \mathcal V_L.
\end{equation}
Moreover, following an integration by parts with respect to the space variable, one obtains:
\begin{equation*}
\label{eq:lag_hwp}
\myint_{\Imega}{ \myint_{\Omega}{ \qs \left( \rho A \qpp - EA \frac{\partial^{2} q}{\partial x^{2}} - f \right) dx } dt}
+ \myint_{\Imega}{ \qs(\ell, t) \left(EA \frac{\partial q}{\partial x} (\ell,t) - g(t) \right) dt }
= 0, \quad \forall \qs \in \mathcal V_L,
\end{equation*}
which allows us to recover the strong form of the wave equation~\eqref{eq:wave_equation} and the Neumann boundary condition~\eqref{eq:NeumannBC}.

\subsubsection{Discrete formulation} 
\label{subsub:lag_fem}


The objective here is to define the discrete problem using a Finite Element method in space and a finite difference approach in time. In order to do so, we consider the semi-weak formulation instead of the weak formulation~\eqref{eq:weak-lagrange}:
\begin{align*}
&\text{Find $q(\cdot,t) \in V$, $\forall t \in \Imega$, such that} \nonumber \\
&\qquad\quad 
\myint_{\Omega}{ \rho(x) A \qs(x) \qpp(x,t) + EA \frac{\partial \qs}{\partial x}(x) \frac{\partial q}{\partial x}(x,t) \, dx} \nonumber \\
&\hspace{2.0in} 
= \myint_{\Omega}{ \qs(x) f(x,t) \, dx} + \qs(\ell) g(t), 
\quad \forall \qs \in V, \quad \forall t \in \Imega \\
&\text{and that satisfies the initial conditions} \nonumber \\
&\hspace{2.0in} 
q(x,0) = u_{0}(x), \quad \forall x \in \Omega, \\
&\hspace{2.0in} 
\qp(x,0) = v_{0}(x), \quad \forall x \in \Omega,
\end{align*}
where $V$ is the vector space of functions defined on $\Omega$ as:
\begin{equation*}
V = \left\{ v \in H^{1}(\Omega);\ v(0) = 0 \right\}.
\end{equation*}
We partition the domain into $N_e$ elements $K_e$ such that $\overline{\Omega} = \cup_{e=1}^{N_e} K_e$ and $\text{Int}(K_i) \cap \text{Int}(K_j) = \varnothing$,\break $\forall i,j=1,\ldots,N_e$, $i\neq j$.
We then associate with the mesh a finite element space $V^h \subset V$,\break $\dim V^{h} = \nx$, based on continuous piecewise polynomial functions defined on $\Omega$:
\begin{equation*}
V^h = \{ v_h \in V:\ v_h|_{K_e} \in \mathbb P_k(K_e),\ e=1,\ldots,N_e\}, 
\end{equation*}
where $\mathbb P_k(K_e)$ denotes the space of polynomial functions of degree $k$ on $K_e$. Let $\phi_i$, $i=1,\ldots,\nx$, denote the basis functions of $V^h$, i.e.\ $V^{h} = \text{span}\{\phi_{i}\}$. We then search for finite element solutions in the form:
\begin{equation*}
q_h(x,t) = \sum_{j = 1}^{\nx} q_{j}(t) \phi_{j}(x)
\end{equation*}
where the degrees of freedom $q_j$ depend on time. The Finite Element problem using the Galerkin method thus reads:
\begin{align*}
&\text{Find $q_h(\cdot,t) \in V^h$, $\forall t \in \Imega$, such that} \nonumber \\
&\qquad\quad 
\myint_{\Omega}{ \rho(x) A \phi_i(x) \qpp_h(x,t) + EA \frac{\partial \phi_i}{\partial x}(x) \frac{\partial q_h}{\partial x}(x,t) \, dx} \nonumber \\
&\hspace{2.0in} 
= \myint_{\Omega}{ \phi_i(x) f(x,t) \, dx} + \phi_i(\ell) g(t), 
\quad \forall i=1,\ldots,\nx, \quad \forall t \in \Imega \\
&\text{and that satisfies the initial conditions} \nonumber \\
&\hspace{2.0in} 
q_h(x,0) = u_{0,h}(x), \quad \forall x \in \Omega, \\
&\hspace{2.0in} 
\qp_h(x,0) = v_{0,h}(x), \quad \forall x \in \Omega,
\end{align*}
where $u_{0,h}$ and $v_{0,h}$ are interpolants or projections of $u_0$ and $v_0$ in the space $V^h$. The above problem can be recast in compact form as:
\begin{align}
\label{eq:discr_wave_equation}
M \Qpp(t) + K \Q(t) &= \F(t), \qquad \forall t \in \Imega \\
\Q(0) &= U_{0}, \nonumber \\
\Qp(0) &= V_{0}, \nonumber
\end{align}
where $M$ and $K$ are the global mass and stiffness matrices, respectively, both being symmetric and positive definite:
\begin{equation*}
M_{ij} = \myint_{\Omega}{ \rho A \phi_i \phi_j \, dx}, \qquad
K_{ij} = \myint_{\Omega}{ E A \phi_i' \phi_j' \, dx}, \qquad
\forall i,j=1,\ldots,\nx,
\end{equation*}
$F(t)$ is the loading vector at time $t$:
\begin{equation*}
F_i(t) = \myint_{\Omega}{ \phi_i(x) f(x,t) \, dx} + \phi_i(\ell) g(t), \qquad \forall i=1,\ldots,\nx,
\end{equation*}
Q(t) is the vector of degrees of freedom:
\begin{equation*}
Q(t) = \begin{bmatrix} q_{1}(t) & \ldots & q_{\nx}(t) \end{bmatrix}\tr
\end{equation*}
and $U_0$ and $V_0$ are the initial vectors:
\begin{equation*}
\begin{aligned}
& U_0 = \begin{bmatrix} u_{0,1} & \ldots & u_{0,\nx} \end{bmatrix}\tr, \\
& V_0 = \begin{bmatrix} v_{0,1} & \ldots & v_{0,\nx} \end{bmatrix}\tr. \end{aligned}
\end{equation*}

A classical approach~\cite{estep, nandy} to discretize in time the system of second-order differential equations~\eqref{eq:discr_wave_equation} consists first in rewriting the system as a system of first-order differential equations by introducing the vector of velocities $\W = \Qp$, i.e.\
\begin{align}
\label{eq:cpl_wave_1} 
\Qp(t) - \W(t) = 0, &\qquad \forall t \in \Imega, \\
\label{eq:cpl_wave_2}
M \dot{\W}(t) + K \Q(t) = \F(t), &\qquad \forall t \in \Imega,
\end{align}
and then in applying the Crank-Nicolson scheme (also referred to as the implicit trapezoidal rule) to both~\eqref{eq:cpl_wave_1} and~\eqref{eq:cpl_wave_2}. Dividing the time domain $\Imega$ into $\nt$ subintervals $\Imega^{n} = [t^{n},t^{n+1}]$, $n=1,\ldots,\nt$ of size $\htt=t^{n+1} - t^{n}$, we evaluate $\Q^n$ and $\W^n$, $n=0,\ldots,\nt$, such that $\Q^0 = U_0$ and $\W^0 = V_0$, and:
\begin{align*}
2 ( \Q^{n+1} - \Q^{n} ) - \htt ( \W^{n} + \W^{n+1} ) &= 0, 
& \forall n=0,\ldots,\nt-1,\\
2 M ( \W^{n+1} - \W^{n} ) + \htt K ( \Q^{n} + \Q^{n+1} ) &= \htt ( \F^{n} + \F^{n+1} ),
& \forall n=0,\ldots,\nt-1.
\end{align*}
The above system of equations can be conveniently recast in matrix form as:
\begin{equation}
\label{eq:midpoint_wave}
\renewcommand{\arraystretch}{1.5}
\begin{bmatrix}
\htt K & 2 M \\
2 M & -\htt M
\end{bmatrix} 
\begin{bmatrix} \Q^{n+1} \\ \W^{n+1} \end{bmatrix}
= \begin{bmatrix}
-\htt K & 2 M \\
2 M & \htt M
\end{bmatrix} 
\begin{bmatrix} \Q^{n} \\ \W^{n} \end{bmatrix} 
+ \htt \begin{bmatrix} \F^{n} + \F^{n+1} \\ 0 \end{bmatrix}, 
\quad \forall n=0,\ldots,\nt-1,
\end{equation}
where we have multiplied the first row by matrix $M$. It is worth noting that the scheme is not only implicit and second-order, but preserves the energy of the system over time~\cite{estep, nandy, iavernaro}. We shall compare the scheme to that obtained using the Hamiltonian framework presented below.

\subsection{The Hamiltonian formalism}

We recall that Lagrange described the evolution of a system in terms of the generalized coordinates $q$, and implicitly of its first derivative $\qp$, in the configuration space. Hamilton extended the work of Lagrange in 1834~\cite{hamilton1834} by describing the evolution of the system in the phase space, introducing the generalized coordinates $q$ and their generalized (or conjugate) momenta $p = \rho A \qp$ as independent quantities. In order to do so, he applied a Legendre transform to the Lagrangian with respect to $\qp$ (with $q$ fixed) and thus defined the Hamiltonian functional $\Ham$ as \cite{nasahwp}:
\begin{equation}
\label{eq:ham_legendre}
\Ham(q,p,t) = \int_\Imega p \qp \, dt - \Lag(q,\qp,t).
\end{equation}
While the Lagrangian is written in terms of a difference between the kinetic energy and the potential energy, the Hamiltonian corresponds to the sum of these two energies. In fact, it actually represents the total energy of the system under study in the case of conservative systems, see below.

\subsubsection{Continuous formulation}

The action $\Act$~\eqref{eq:action} is defined in terms of the Hamiltonian functional~\eqref{eq:ham_legendre} as:
\begin{equation*}
\mathcal{S}[q,p] = \myint_{\Imega}{\qp p - \Ham(q,p,t) \, dt}
\end{equation*}
where the Hamiltonian functional for our problem reads:
\begin{equation*}
\Ham(q, p, t) = \underbrace{ \frac{1}{2}\myint_{\Omega}{ \frac{1}{\rho A} p^{2} \, dx} }_{\text{Kinetic Energy}} + \underbrace{ \frac{1}{2}\myint_{\Omega}{EA \left( \frac{\partial q}{\partial x} \right)^{2} \, dx} }_{\text{Potential Energy}} - \underbrace{ \left( \myint_{\Omega}{ f q \, dx} + g(t) q(\ell, t) \right) }_{\text{External Energy}}
\end{equation*}
The generalized coordinate and momenta are searched in the spaces:
\begin{equation*}
\begin{aligned}
&\mathcal U_H = \{ q \in L^2(\Imega;H^1(\Omega));\ q(0,t) = 0,\ \forall t \in \Imega;\ q(x,0) = u_0(x),\ \forall x \in \Omega\}, \\
&\mathcal W_H = \{ p \in L^2(\Imega;L^2(\Omega));\ p(x,0) = \rho A v_0(x),\ \forall x \in \Omega\}.
\end{aligned}
\end{equation*}
The Hamilton's Weak Principle then states that the trajectory $(q,p)$ of the system in the phase space should satisfy:
\begin{equation*}
\Act'[q,p](\qs,\ps) = \bigg[ \myint_{\Omega}{ \qs p  \, dx }\bigg]_{0}^{T},
\end{equation*}
where $\Act'[q,p](\qs,\ps)$ denotes here the G\^ateaux derivative of $\Act[q,p]$ with respect to a perturbation $(\qs,\ps)$ belonging to the spaces:
\begin{equation*}
\begin{aligned}
&\mathcal V_\text{H} = \{ \qs \in L^2(\Imega;H^1(\Omega));\ \qs(0,t) = 0,\ \forall t \in \Imega;\ \qs(x,0) = 0,\ \forall x \in \Omega\}, \\
&\mathcal Z_\text{H} = \{ \ps \in L^2(\Imega;L^2(\Omega));\ \ps(x,0) = 0,\ \forall x \in \Omega\}.
\end{aligned}
\end{equation*}
We easily compute the G\^ateaux derivative of $\Act[q,p]$ as:
\[
\Act'[q,p](\qs,\ps) = 
\myint_{\Imega}{\myint_{\Omega}{\qsp p + \ps\qp - \frac{1}{\rho A} \ps p - EA \frac{\partial \qs}{\partial x} \frac{\partial q}{\partial x} + \qs f\, dx}dt}\
+ \myint_{\Imega}{\qs(\ell, t) g(t) \, dt},
\]
so that a weak formulation of the problem reads:
\begin{equation}
\label{eq:weak-Hamilton}
\begin{aligned}
&\text{Find $(q,p) \in \mathcal U_H \times \mathcal W_H$ such that}\\
&\qquad\quad 
\myint_{\Imega}{\myint_{\Omega}{\qsp p + \ps\qp - \frac{1}{\rho A} \ps p - EA \frac{\partial \qs}{\partial x} \frac{\partial q}{\partial x} dx} dt}\ 
- \myint_{\Omega}{\qs(T) p(T)\, dx} \\
&\hspace{1.5in} 
= - \myint_{\Imega}{\myint_{\Omega}{\qs f\, dx} dt}\ 
- \myint_{\Imega}{\qs(\ell, t) g(t) \, dt}, 
\qquad \forall (\qs, \ps) \in \mathcal V_H \times \mathcal Z_H.
\end{aligned}
\end{equation}
Integrating by parts with respect to time and space for sufficiently smooth data, we can rewrite~\eqref{eq:weak-Hamilton} as:
\begin{equation*}
\label{eq:ham_hwp}
\begin{aligned}
&\myint_{\Imega}{\myint_{\Omega}{ \ps \left( \frac{1}{\rho A} p - \qp \right) 
+ \qs \left( \ppoint - EA \frac{\partial^2 q}{\partial x^2} - f \right) dx } dt}\ \\
&\hspace{1.5in}
+ \myint_{\Imega}{ \qs(\ell, t) \left(EA \frac{\partial q}{\partial x} (\ell,t) - g(t) \right) dt}
= 0, \qquad \forall (\qs, \ps) \in \mathcal V_H \times \mathcal Z_H,
\end{aligned}
\end{equation*}
or, in a decoupled fashion with respect to the test functions $\qs$ and $\ps$, as:
\begin{align*}
& \myint_{\Imega}{\myint_{\Omega}{\ps \left( \frac{1}{\rho A} p - \qp \right) dx } dt} = 0, 
& \forall \ps \in \mathcal Z_H, \\
& \myint_{\Imega}{\myint_{\Omega}{\qs \left( \ppoint - EA \frac{\partial^{2} q}{\partial x^{2}} - f \right) dx } dt}\ 
+ \myint_{\Imega}{ \qs(\ell, t) \left(EA \frac{\partial q}{\partial x} (\ell,t) - g(t) \right) dt} =0, 
& \forall \qs \in \mathcal V_H.
\end{align*}
The Hamiltonian formulation (\ref{eq:weak-Hamilton}) can be viewed as a mixed problem for which the coupled differential equations in strong form are given by:
\begin{equation}
\label{eq:can_ham}
\begin{aligned}
& \qp = \dfrac{1}{\rho A} p, 
& \quad \forall x \in \Omega, \forall t \in \Imega, \\
& \ppoint = EA \dfrac{\partial^{2} q}{\partial x^{2}} + f, 
& \quad \forall x \in \Omega, \forall t \in \Imega.
\end{aligned}
\end{equation}
We observe that the system of equations is equivalent to~\eqref{eq:wave_equation} by introducing the auxiliary variable $p=\rho A \qp$. The system of equations~\eqref{eq:can_ham} is usually referred to as the canonical Hamilton equations. One advantage of the Hamiltonian formalism is that it explicitly informs one on how to define this auxiliary variable.

\subsubsection{Discrete formulation} 
\label{subsub:ham_fem}

For convenience, we first recast the weak formulation~\eqref{eq:weak-Hamilton} as the system of equations:
\begin{alignat}{3}
\label{eq:weak-Ham1}
& \myint_{\Imega}{ \myint_{\Omega}{ \ps \left( \frac{1}{\rho A} p - \qp \right) dx } dt} = 0, 
& \qquad \forall \ps \in \mathcal Z_H, \\
\label{eq:weak-Ham2}
& \myint_{\Imega}{ \myint_{\Omega}{ \qs \ppoint + E A \frac{\partial \qs}{\partial x} \frac{\partial q}{\partial x}\, dx} dt} 
= \myint_{\Imega}{ \myint_{\Omega}{ \qs f \, dx } dt}\ 
+ \myint_{\Imega}{ \qs(\ell, t) g(t) \, dt},
& \qquad \forall \qs \in \mathcal V_H.
\end{alignat}

The objective here is to discretize the set of equations using a Finite Element method in both space and time. For the spatial discretization, we emply the same mesh and FE space $V^h$ for $q$ and $p$ as the ones used in the Lagrangian formulation. In other words, we search for Finite Element solutions in the form:
\[
\begin{aligned}
& q_h(x,t) = \sum_{j=1}^{\nx} q_j(t) \phi_j(x), \\
& p_h(x,t) = \sum_{j=1}^{\nx} p_j(t) \phi_j(x), 
\end{aligned}
\]
and denote by $\Q$ and $\PP$ the vectors of time-dependent degrees of freedom $q_j$ and $p_j$, $j=1,\ldots,\nx$, respectively. In the same manner, we consider test functions in the form:
\[
\begin{aligned}
& \qs_h(x,t) = \sum_{i=1}^{\nx} \qs_i(t) \phi_i(x), \\
& \ps_h(x,t) = \sum_{i=1}^{\nx} \ps_i(t) \phi_i(x). 
\end{aligned}
\]
Inserting the trial and test functions in~\eqref{eq:weak-Ham1} and~\eqref{eq:weak-Ham2}, one obtains the semi-discrete set of equations:
\begin{align}
\label{eq:cpl_ham_1}
& \myint_{\Imega}{\Ps{\tr}(t) \left( \Mbb\PP(t) - \Mb\Qp(t) \right) dt} = 0, \\
\label{eq:cpl_ham_2}
& \myint_{\Imega}{\Qs{\tr}(t) \left( \Mb\Pp(t) + K\Q(t) \right) dt} = \myint_{\Imega}{\Qs{\tr}(t)\F(t)\, dt},
\end{align}
where the matrices $\Mb$ and $\Mbb$ are the symmetric positive-definite matrices:
\begin{equation*}
\Mb_{ij} = \myint_{\Omega} \phi_i \phi_j \, dx, \qquad
\Mbb_{ij} = \myint_{\Omega} \frac{1}{\rho A} \phi_i \phi_j \, dx, 
\qquad \forall i,j=1,\ldots,\nx.
\end{equation*}

In order to approximate the functions $\Q$ and $\PP$ with respect to time, we follow an approach similar to the one proposed in~\cite{nasahwp}. We thus consider continuous piecewise linear trial functions on each subinterval $\Imega^{n} = [t^{n},t^{n+1}]$, $n=0,\ldots,\nt-1$:
\begin{equation*}
\begin{aligned}
&\Q(t) \approx 
\left[ \frac{t^{n+1} - t}{\htt} \right] \Q^{n} + 
\left[ \frac{t - t^{n}}{\htt} \right] \Q^{n+1}\\
&\PP(t) \approx 
\left[ \frac{t^{n+1} - t}{\htt} \right] \PP^{n} + 
\left[ \frac{t - t^{n}}{\htt} \right] \PP^{n+1}\\
\end{aligned}
\end{equation*}
and piecewise constant test functions on each subinterval $\Imega^n$, i.e.
\begin{equation*}
\Qs(t) = {\Qs}^n, \qquad \Ps(t) = {\Ps}^n.
\end{equation*}
Using the above expressions in~\eqref{eq:cpl_ham_1} and~\eqref{eq:cpl_ham_2} gives:
\begin{align*}
&\midsum_{n=0}^{\nt-1}
\left[{\Ps}^{n}\right]\tr 
\myint_{\Imega^n} 
\Mbb \left( 
\left[ \frac{t^{n+1} - t}{\htt} \right] \PP^{n} + 
\left[ \frac{t - t^{n}}{\htt}   \right] \PP^{n+1} \right) - 
\Mb \left( \frac{\Q^{n+1} - \Q^{n}}{\htt} \right) dt = 0, \\
&\midsum_{n=0}^{\nt - 1} 
\left[{\Qs}^{n}\right]\tr 
\myint_{\Imega^{n}}
\Mb \left( \frac{\PP^{n+1} - \PP^{n}}{\htt} \right) + 
K \left( 
\left[ \frac{t^{n+1} - t}{\htt} \right] \Q^{n} + 
\left[ \frac{t - t^{n}}{\htt}   \right] \Q^{n+1} \right) dt \\
&\hspace{2.0in} 
= \midsum_{n=0}^{\nt - 1}
\left[{\Qs}^{n}\right]\tr
\myint_{\Imega^{n}} 
\left[ \frac{t^{n+1} - t}{\htt} \right] \F^{n} + 
\left[ \frac{t - t^{n}}{\htt}   \right] \F^{n+1} dt,
\end{align*}
which, after integration, yields:
\begin{align*}
&\midsum_{n=0}^{\nt - 1}
\left[{\Ps}^{n}\right]\tr 
\left[ 
\frac{\htt}{2} \Mbb \Big( \PP^{n} + \PP^{n+1} \Big) - 
\Mb \Big( \Q^{n+1} - \Q^{n} \Big) \right] = 0, \\
&\midsum_{n=0}^{\nt - 1}
\left[{\Qs}^{n}\right]\tr 
\left[ 
\Mb \Big( \PP^{n+1} - \PP^{n} \Big) + 
\frac{\htt}{2} K \Big( \Q^{n} + \Q^{n+1} \Big) - 
\frac{\htt}{2} \Big( \F^{n} + \F^{n+1} \Big) \right]
= 0.
\end{align*}
Since the above equations hold for any arbitrary test fields, the problem consists in solving for $\Q^n$ and $\PP^n$, $n=0,\ldots,\nt$, such that $\Q^0 = U_0$ and $\PP^0 = \rho A V_0$, and:
\begin{align*}
& 2 \Mb \Q^{n+1} - \htt \Mbb \PP^{n+1} 
= 2 \Mb \Q^{n}  + \htt \Mbb \PP^{n},
&\forall n=0,\ldots,\nt-1,\\
& \htt K \Q^{n+1} + 2 \Mb \PP^{n+1} 
= - \htt K \Q^{n} + 2 \Mb \PP^{n} 
+ \htt \Big( \F^{n} + \F^{n+1} \Big),
&\forall n=0,\ldots,\nt-1,
\end{align*}
which can be conveniently recast in matrix form as:
\begin{equation}
\label{eq:midpoint_ham}
\renewcommand{\arraystretch}{1.5}
\begin{bmatrix}
\htt K & 2 \Mb \\
2 \Mb & - \htt \Mbb
\end{bmatrix} \begin{bmatrix}
\Q^{n+1} \\ \PP^{n+1}
\end{bmatrix} = \begin{bmatrix}
- \htt K & 2 \Mb \\
2 \Mb & \htt \Mbb
\end{bmatrix} \begin{bmatrix}
\Q^{n} \\ \PP^{n}
\end{bmatrix} + \htt \begin{bmatrix}
\F^{n} + \F^{n+1} \\ 0
\end{bmatrix},
\quad \forall n=0,\ldots,\nt-1.
\end{equation}
We note that we would have arrived exactly at the same set of equations if we had chosen to discretize the problem in time by the finite differences Crank-Nicolson scheme. More interestingly, we observe that~\eqref{eq:midpoint_ham} has exactly the same structure as~\eqref{eq:midpoint_wave} except for the fact that matrix $M$ in~\eqref{eq:midpoint_wave} has been replaced by either $\Mb$ or $\Mbb$. We shall see in the following how this slight difference will affect the results within the PGD framework.
 
Finally, we can conveniently collect the degrees of freedom of the FE solution into the matrix $U$ of size $\nx \times \nt$:
\begin{align*}
U &= \begin{bmatrix}
\Q^{1} & \Q^{2} & \ldots & \Q^{\nt}
\end{bmatrix} = \begin{bmatrix}
q_{1}^{1} & q_{1}^{2} & \ldots & q_{1}^{\nt} \\
q_{2}^{1} & q_{2}^{2} & \ddots & \vdots \\
\vdots & \ddots & \ddots & \vdots \\
q_{\nx}^{1} & \ldots & \ldots & q_{\nx}^{\nt}
\end{bmatrix}.
\end{align*}
We will use the solutions given by (\ref{eq:midpoint_wave}) and (\ref{eq:midpoint_ham}) as reference solutions when assessing the results of the PGD. In particular, we will perform some Singular Value Decomposition of $U$ to identify the principal components or modes of the FE solution.

\section{PGD reduced-order modeling}
\label{sect:PGD}

The PGD framework aims at searching for an approximation of the given field of interest, e.g.\ the generalized coordinate $q$, in the separated form:
\begin{equation*}
\label{eq:PGDrepresentation}
q(x,t) \approx q_{m}(x,t) 
= q_{0}(x,t) + \sum_{i=1}^{m} \lambda_{i}(t)\mu_{i}(x),
\end{equation*}
where the truncation parameter $m$ denotes the number of modes in the representation, the $\lambda_{i}$'s and the $\mu_{i}$'s stand for the temporal and spatial modes, respectively. $q_{0}$ is a lift function that satisfies the non-homogeneous Dirichlet boundary conditions and initial conditions so that each mode in the separated representation satisfies the corresponding homogeneous boundary and initial conditions~\cite{chinesta}.

The PGD solution is often computed based on a greedy algorithm~\cite{boucinha2013, chinesta}. Assuming that the $q_{m-1}$ mode has been computed, the approach consists then in finding the $m\textsuperscript{th}$ enrichment mode as follows:
\begin{equation*}
q_{m}(x,t) = q_{m-1}(x,t) + \lambda(t)\mu(x),
\end{equation*}
where the subscript $m$ has been dropped from $\lambda_m$ and $\mu_m$ for the sake of clarity in the notation. Inserting the trial solution $q_m$ in the governing differential equations using the Galerkin method or a residual minimization approach~\cite{boucinha2013, chinesta, billaud} leads to the solution of a non-linear system for the unknown functions $\lambda$ and $\mu$. The problem is usually solved by means of an appropriate iterative scheme, such as a fixed point algorithm that will be considered here, in which one determines in an alternating fashion at each iteration of the algorithm the solution $\mu$ with $\lambda$ known and then $\lambda$ with $\mu$ known~\cite{chinesta, cognard}.

We describe below the derivation of the PGD formulation using the Lagrangian and Hamiltonian formalism. We will construct first the formulations at the continuous level and then propose some numerical schemes to discretize the problems.

\subsection{Lagrangian-based PGD}

We consider first the Lagrangian framework. We start from the formulation~\eqref{eq:pre_lag_hwp} as we will use a Finite Element approach in space and a Finite Differences approach in time.

\subsubsection{Continuous formulation}

Substituting $q_{m-1}(x, t) + \lambda(t)\mu(x)$ for $q_{m}(x, t)$ in~\eqref{eq:pre_lag_hwp}, one gets:
\begin{equation}
\label{eq:lag_pgd}
\begin{aligned}
&\myint_{\Imega} \myint_{\Omega} \rho A \qs \lambdapp \mu + EA \frac{\partial \qs}{\partial x} \lambda \mu' \, dxdt \\
&\qquad\qquad 
= \myint_{\Imega}{ \myint_{\Omega}{ \qs ( f - \rho A \ddot{q}_{m-1} ) - EA \frac{\partial \qs}{\partial x} \frac{\partial q_{m-1}}{\partial x} \, dx } dt} 
+ \myint_{\Imega}{\qs(\ell,t) g(t) \, dt}, \qquad \forall \qs \in \mathcal V_L,
\end{aligned}
\end{equation}
where the vector space $\mathcal V_L$ is defined as in~\eqref{eq:VLagrange}. In view of using a fixed point approach, we now derive the problems for $\mu$ and for $\lambda$.

We assume first that $\lambda$ is known and search for $\mu \in V$. We thus choose test functions in the form $\qs(x,t) = \lambda(t) \mus(x)$ with $\mus \in V$. Equation~\eqref{eq:lag_pgd} thus reduces to:
\begin{equation}
\label{eq:space_pgd_lag}
\myint_{\Omega} m_{\ell t} \mu \mus + k_{\ell t} \mu' {\mus}' dx 
= \myint_{\Omega}{ r_{\ell \mu} (\mus) \, dx} 
+ \left( \myint_{\Imega}{ \lambda g(t) \, dt} \right) \mus(\ell), \qquad \forall \mus \in V,
\end{equation}
with:
\begin{align*}
& m_{\ell t} = \rho A \myint_{\Imega}{ \lambdapp \lambda \, dt} = \rho A \left( \lambdap(T) \lambda(T) - \myint_{\Imega}{ \lambdap^{2} \ dt } \right), \\
& k_{\ell t} = EA \myint_{\Imega}{ \lambda^{2} \, dt }, \\
& r_{\ell \mu}(\mus) 
= \left( \myint_{\Imega} f\lambda \, dt \right) \mus 
- \rho A \left( \myint_{\Imega} \qpp_{m-1}\lambda \, dt \right) \mus
- EA \left( \myint_{\Imega} \frac{\partial q_{m-1}}{\partial x} \lambda \, dt  \right) {\mus}',
\end{align*}
where $m_{\ell t}$, $k_{\ell t}$, and $r_{\ell \mu}(\mus)$ are possibly functions of the spatial variable only.

Similarly, we assume now that $\mu$ is known and search for $\lambda=\lambda(t)$. Choosing test functions in the form $\qs(x,t) = \mu(x) \lambdas(t)$, Equation~\eqref{eq:lag_pgd} then becomes:
\begin{equation}
\label{eq:time_pgd_lag}
\myint_{\Imega}{ \lambdas \left( m_{\ell x} \lambdapp + k_{\ell x} \lambda \right) \, } dt = \myint_{\Imega}{ \lambdas r_{\ell \lambda} \, dt }, \qquad \forall \lambdas, 
\end{equation}
or simply:
\begin{equation}
m_{\ell x} \lambdapp(t) + k_{\ell x} \lambda(t) = r_{\ell \lambda}(t), \qquad \forall t \in \Imega,
\end{equation}
with:
\begin{align*}
&m_{\ell x} = \myint_{\Omega}{ \rho A \mu^{2} \, dx}, \\
&k_{\ell x} = \myint_{\Omega}{ EA \left( \mu' \right)^{2} dx }, \\
&r_{\ell \lambda}(t)  
= \myint_{\Omega} f\mu \, dx 
- \myint_{\Omega}  \rho A \qpp_{m-1} \mu\, dx 
- \myint_{\Omega} EA \frac{\partial q_{m-1}}{\partial x} \mu' \, dx 
+ \mu(\ell) g(t),
\end{align*}
where $m_{\ell x}$ and $k_{\ell x}$ are constant.

\subsubsection{Discrete formulation}

As before, we discretize~\eqref{eq:space_pgd_lag} by the Finite Element method. In other words, we are looking for an approximate solution $\mu_h \in V^h$ of $\mu$:
\begin{equation*}
\mu(x) \approx \mu_h(x) = \sum_{j=1}^{\nx} \mu_j \phi_j(x),
\end{equation*}
satisfying:
\begin{equation*}
\myint_{\Omega} m_{\ell t} \mu_h \phi_i + k_{\ell t} \mu_h' \phi_i' dx 
= \myint_{\Omega}{ r_{\ell \mu} (\phi_i) \, dx} 
+ \left( \myint_{\Imega}{ \lambda g(t) \, dt} \right) \phi_i(\ell), \qquad \forall i=1,\ldots,\nx,
\end{equation*}
which can be recast in matrix form as:
\begin{equation}
\label{eq:disc_slag}
\Big(  M_{\ell t} + K_{\ell t} \Big) U_{\ell\mu} = R_{\ell\mu} .
\end{equation}
Here, the matrices $M_{\ell t}$ and $K_{\ell t}$ are the modified mass and stiffness matrices, respectively: 
\begin{align*}
M_{\ell t} = M \left( \myint_{\Imega}{ \lambdapp \lambda \ dt} \right),
\qquad
K_{\ell t} = K \left( \myint_{\Imega}{ \lambda^{2} \ dt } \right),
\end{align*}
and the vector of degrees of freedom $U_{\ell\mu}$ and the loading vector $R_{\ell\mu}$ are given by:
\[
U_{\ell\mu} = \begin{bmatrix} \vdots \\ \mu_i \\ \vdots \end{bmatrix},
\qquad
R_{\ell\mu} = 
\begin{bmatrix}
\vdots \\ \myint_{\Omega}{ r_{\ell \mu} (\phi_i) \, dx} + \left( \myint_{\Imega}{ \lambda g(t) \, dt} \right) \phi_i(\ell) \\ \vdots 
\end{bmatrix}.
\]

For the time discretization, we again reduce the second-order equation~\eqref{eq:time_pgd_lag} to a system of first-order equations by introducing the new variable $\omega = \lambdap$, i.e.
\begin{align*}
\lambdap(t) - \omega(t) = 0, \qquad \forall t \in \Imega, \\
m_{\ell x} \dot{\omega}(t) +  k_{\ell x} \lambda(t) = r_{\ell \lambda}(t), \qquad \forall t \in \Imega.
\end{align*}
and apply the Crank-Nicolson scheme to each equation. The scheme consists then in finding the pair $(\lambda_n,\omega_n) \in \mathbb R^2$, $n=1,\ldots,\nt$ such that:
\begin{equation}
\label{eq:midpoint_pgd_lag}
\renewcommand{\arraystretch}{1.5}
\begin{bmatrix} \htt k_{\ell x} & 2 m_{\ell x} \\ 2 m_{\ell x} & -\htt m_{\ell x} \end{bmatrix} 
\begin{bmatrix} \lambda^{n+1} \\ \omega^{n+1} \end{bmatrix} = 
\begin{bmatrix} -\htt k_{\ell x} & 2 m_{\ell x} \\ 2 m_{\ell x} & \htt m_{\ell x} \end{bmatrix} 
\begin{bmatrix} \lambda^{n} \\ \omega^{n} \end{bmatrix} + \htt 
\begin{bmatrix} r_{\ell \lambda}^{n} + r_{\ell \lambda}^{n+1} \\ 0 \end{bmatrix}, 
\quad \forall n=0,\ldots, \nt-1,
\end{equation}
where we have multiplied the first row by $m_{\ell x}$.
We observe that the above system of equations has naturally the same structure as that in~\eqref{eq:midpoint_wave}.

\subsection{Hamiltonian-based PGD}

The proper-generalized decomposition method applied within the Hamiltonian framework aims at approximating both the 
generalized coordinates $q$ and their generalized momenta $p$ in the separated form:
\begin{align*}
&q(x,t) \approx q_{m}(x,t) = q_{m-1}(x,t) + \lambda(t)\mu(x), \\
&p(x,t) \approx p_{m}(x,t) = p_{m-1}(x,t) + \omega(t) \nu(x).
\end{align*}
The goal in this section is to construct the problems that satisfy the enrichment modes $\lambda(t)\mu(x)$ and $\omega(t) \nu(x)$ assuming that both $q_{m-1}(x,t)$ and $p_{m-1}(x,t)$ have been calculated. We present first the continuous formulation of the problems. 

\subsubsection{Continuous formulation}

Replacing $q$ and $p$ in~\eqref{eq:weak-Hamilton} by $q_m$ and $p_m$, respectively, one straightforwardly gets:
\begin{alignat}{2}
\label{eq:ham_pgd1} 
&\myint_{\Imega}{ \myint_{\Omega}{ \ps \left( \frac{1}{\rho A} \omega \nu - \lambdap \mu \right) dx } dt} 
= - \myint_{\Imega}{ \myint_{\Omega}{ \ps \left( \frac{1}{\rho A} p_{m-1} - \dot{q}_{m-1} \right) dx } dt}, 
& \qquad \forall \ps \in \mathcal Z_H, \\
\nonumber
&\myint_{\Imega}{ \myint_{\Omega}{ \qs \dot{\omega}\nu + EA \frac{\partial \qs}{\partial x} \lambda \mu' \, dx } dt} 
= \myint_{\Imega}{ \myint_{\Omega}{ \qs ( f - \dot{p}_{m-1} ) - EA \frac{\partial \qs}{\partial x} \frac{\partial q_{m-1}}{\partial x} \, dx } dt} \\
\label{eq:ham_pgd2}
&\hspace{3.5in}+ \myint_{\Imega}{ \qs(\ell, t) g(t)\, dt}, 
&\qquad \forall \qs \in \mathcal V_H.
\end{alignat}

As in the Lagrangian procedure, we first assume that $\lambda$ and $\omega$ are known and search for the solutions $\mu \in V$ and $\nu \in L^2(\Omega)$. We therefore choose test functions in the form $\qs(x,t) = \lambda(t) \mus(x)$ and $\ps(x, t) = \omega(t) \nus(x)$. Equations~\eqref{eq:ham_pgd1} and~\eqref{eq:ham_pgd2} thus become:
\begin{equation}
\label{eq:space_pgd_ham}
\begin{aligned}
& \myint_{\Omega} m_{ht}\nu\nus - c_{ht}\mu\nus \, dx 
= \myint_{\Omega} r_{h\nu} (\nus) \, dx, 
&&\quad \forall \nus \in L^2(\Omega), \\
& \myint_{\Omega} d_{ht} \nu \mus + k_{ht} \mu' {\mus}' dx 
= \myint_{\Omega} r_{h \mu}(\mus) \, dx 
+ \left(\myint_{\Imega} \lambda g(t) \, dt\right) \mus(\ell),
&&\quad \forall \mus \in V,
\end{aligned}
\end{equation}
with:
\begin{align*}
&m_{ht} = \frac{1}{\rho A} \myint_{\Imega} \omega^{2}\, dt, \\
&c_{ht} = \myint_{\Imega} \lambdap\omega  \, dt, \\
&d_{ht} = \myint_{\Imega} \dot{\omega} \lambda \, dt = \nu(T) \lambda(T) - c_{ht}, \\
&k_{ht} = EA \myint_{\Imega} \lambda^{2} \, dt, \\
&r_{h\nu}(\nus) 
= \left( \myint_{\Imega} \qp_{m-1} \omega \, dt \right) \nus
- \dfrac{1}{\rho A} \left( \myint_{\Imega} p_{m-1} \omega \, dt \right) \nus, \\
&r_{h\mu}(\mus) 
= \left( \myint_{\Imega} f \lambda \, dt \right) \mus 
- \left( \myint_{\Imega} \ppoint_{m-1} \lambda \, dt \right) \mus 
- EA \left( \myint_{\Imega} \dfrac{\partial q_{m-1}}{\partial x} \lambda \, dt \right) {\mus}' .
\end{align*}
We remark that $c_{ht}$ and $d_{ht}$ are constant while $m_{ht}$ and $k_{ht}$ could possibly depend on the space variable. 

To construct the problem in time, we suppose that $\mu$ and $\nu$ are known and look for $\lambda$ and $\omega$. Choosing the tests functions as $\qs(x,t) = \mu(x)\lambdas(t)$ and $\ps(x,t) = \nu(x) \omegas(t)$ in~\eqref{eq:ham_pgd1} and~\eqref{eq:ham_pgd2}, one obtains:
\begin{equation}
\label{eq:time_pgd_ham}
\begin{aligned}
& \myint_{\Imega} c_{hx} \lambdap\omegas - m_{hx} \omega\omegas \, dt 
= \myint_{\Imega} r_{h\omega} (\omegas) \, dt,
&&\quad \forall \omegas, \\
& \myint_{\Imega} c_{hx} \dot{\omega} \lambdas + k_{hx} \lambda\lambdas \, dt 
= \myint_{\Imega} r_{h \lambda} (\lambdas) \, dt 
+ \left( \myint_{\Imega} \lambdas g(t) \, dt \right) \mu(\ell),
&&\quad \forall \lambdas,
\end{aligned}
\end{equation}
with:
\begin{align*}
&m_{hx} = \myint_{\Omega} \frac{1}{\rho A} \nu^{2} \, dx \\
&c_{hx} = \myint_{\Omega} \mu \nu \, dx \\
&k_{hx} = \myint_{\Omega} EA (\mu')^{2} \, dx \\
&r_{h\omega}(\omegas) 
= \left( \myint_{\Omega} \dfrac{1}{\rho A} p_{m-1} \nu \, dx \right) \omegas
- \left( \myint_{\Omega} \qp_{m-1}\nu \, dx \right) \omegas \\
&r_{h\lambda}(\lambdas) 
= \left( \myint_{\Omega} f\mu \, dx \right) \lambdas
- \left( \myint_{\Omega} \ppoint_{m-1} \mu \, dx \right) \lambdas
- \left( \myint_{\Omega} EA \dfrac{\partial q_{m-1}}{\partial x} \mu' \, dx \right) \lambdas
\end{align*}

\subsubsection{Discrete formulation}

For the discretization of Problem~(\ref{eq:space_pgd_ham}) in space, we are looking for finite element solutions $\mu_h \in V^h$ and $\nu_h \in V^h$ of $\mu$ and $\nu$, respectively, such that:
\begin{equation*}
\begin{aligned}
&\mu(x) \approx \mu_h(x) = \sum_{j=1}^{\nx} \mu_j \phi_j(x),\\
&\nu(x) \approx \nu_h(x) = \sum_{j=1}^{\nx} \nu_j \phi_j(x),
\end{aligned}
\end{equation*}
that satisfy the system of coupled equations:
\begin{equation*}
\begin{aligned}
& \myint_{\Omega} m_{ht}\nu_h\phi_i - c_{ht}\mu_h\phi_i \, dx 
= \myint_{\Omega} r_{h\nu} (\phi_i) \, dx, 
&&\quad \forall i=1,\ldots, \nx,\\
& \myint_{\Omega} d_{ht} \nu_h \phi_i + k_{ht} \mu_h' \phi_i' dx 
= \myint_{\Omega} r_{h \mu}(\phi_i) \, dx 
+ \left(\myint_{\Imega} \lambda g(t) \, dt\right) \phi_i(\ell),
&&\quad \forall i=1,\ldots, \nx,
\end{aligned}
\end{equation*}
which can be equivalently written as:
\begin{equation}
\label{eq:disc_sham}
\renewcommand{\arraystretch}{1.5}
\begin{bmatrix} \phantom{-} K_{h t} & D_{h t} \\ - C_{h t}  & M_{h t} \end{bmatrix}
\begin{bmatrix} U_{h\mu} \\ U_{h\nu} \end{bmatrix} 
= 
\begin{bmatrix} R_{h\mu} \\ R_{h\nu} \end{bmatrix}.
\end{equation}
The stiffness and mass matrices are given here as:
\begin{equation*}
\begin{aligned}
&K_{h t} = K \left( \myint_{\Imega}{ \lambda^{2} \, dt } \right), 
&&\qquad D_{h t} = \Mb \left( \myint_{\Imega}{ \dot{\omega} \lambda \, dt} \right), \\
&C_{h t} = \Mb \left( \myint_{\Imega}{ \omega \lambdap \, dt} \right),
&&\qquad M_{h t} = \Mbb\left( \myint_{\Imega}{ \omega^{2} \, dt} \right),
\end{aligned}
\end{equation*}
while the vectors $U_{h\mu}$ and $U_{h\nu}$ are the vectors of the degrees of freedom associated with $\mu_h$ and $\nu_h$, respectively, and the loading vectors $R_{h\mu}$ and  $R_{h\nu}$ are the residual vectors corresponding to $r_{h\mu}$ and $r_{h\nu}$.

Proceeding as before, the discretization in time of the system of equations~\eqref{eq:time_pgd_ham} using the algorithm described in Section~\ref{subsub:ham_fem} leads to:
\begin{equation}
\label{eq:midpoint_pgd_ham}
\renewcommand{\arraystretch}{1.5}
\begin{bmatrix} \htt k_{h x} & 2 c_{h x} \\ 2 c_{h x} & - \htt m_{h x} \end{bmatrix} 
\begin{bmatrix} \lambda^{n+1} \\ \omega^{n+1} \end{bmatrix} = 
\begin{bmatrix} - \htt k_{h x} & 2 c_{h x} \\ 2 c_{h x} & \htt m_{h x} \end{bmatrix} 
\begin{bmatrix} \lambda^{n} \\ \omega^{n} \end{bmatrix} + \htt 
\begin{bmatrix} r_{h \lambda}^{n} + r_{h \lambda}^{n+1} \\ r_{h\omega}^{n} + r_{h\omega}^{n+1} \end{bmatrix},
\quad \forall n=0,\ldots,\nt-1.
\end{equation}
We observe that the above system is slightly different from the one obtained by the Lagrangian approach~\eqref{eq:midpoint_pgd_lag}.

\subsection{Updating procedure of the temporal modes and Gram-Schmidt process} 
\label{sub:update}

Let us consider a separable function $q(x,t)$. At the $m$\textsuperscript{th} enrichment, the PGD approximation of $q$ is given by:
\begin{equation*}
q(x,t) \approx q_{m}(x,t) 
= q_{0}(x,t) + \sum_{i=1}^{m}{ \lambda_{i}(t)\mu_{i}(x) } 
= q_{m-1}(x, t) + \lambda_{m}(t)\mu_{m}(x).
\end{equation*}
The algorithm used here to construct the modes is based on a greedy approach; namely, each enrichment step aims at the determination of the new mode $(\mu_{m}, \lambda_{m})$. The pair $(\mu_{m}, \lambda_{m})$ is thus computed based on the information contained in $(\mu_{k}, \lambda_{k})_{1 \leqslant k \leqslant m-1}$. However, the previously computed modes $(\mu_{k}, \lambda_{k})_{1 \leqslant k \leqslant m-1}$ do not benefit from the new information introduced by $(\mu_{m}, \lambda_{m})$.

One idea to improve the convergence of the PGD approximation is to update the temporal modes in a global manner~\cite{nouy, alameddin, boisse, bonithon}. In other words, after a new mode is found, the updating algorithm reevaluates the modes $(\lambda_{k})_{1 \leqslant k \leqslant m}$ in order to obtain a better combination of the spatial modes $(\mu_{k})_{1 \leqslant k \leqslant m}$.

This procedure will significantly improve the convergence of the PGD at a fairly low computational cost, which only depends on $\nt$ and $m$~\cite{alameddin}. However, this procedure requires the computation of some matrices that can become ill-conditioned with the increase of the number of modes. If the later occurs, the procedure causes instabilities.

\subsubsection{Lagrangian update} 
\label{subsub:lag_update}

We consider~\eqref{eq:pre_lag_hwp} and solve for $(\lambda_{k})_{1 \leqslant k \leqslant m}$, with $(\mu_{k})_{1 \leqslant k \leqslant m}$ known, using the following trial and test functions:
\begin{equation*}
q_m(x,t) = q_{0}(x,t) + \sum_{i=1}^{m} \lambda_{i}(t)\mu_{i}(x), \qquad 
\qs(x,t) = \sum_{i=1}^{m} \lambdas_{i}(t)\mu_{i}(x).
\end{equation*}
After discretization, we obtain the following system of equations:
\begin{equation}
\label{eq:midpoint_pgd_lag_update}
\renewcommand{\arraystretch}{1.5}
\begin{bmatrix} \htt K_{\ell x} & 2 M_{\ell x} \\ 2 \Mb_{\ell x} & -\htt \Mb_{\ell x} \end{bmatrix} 
\begin{bmatrix} \llambda^{n+1} \\ \oomega^{n+1} \end{bmatrix} = 
\begin{bmatrix} -\htt K_{\ell x} & 2 M_{\ell x} \\ 2 \Mb_{\ell x} & \htt \Mb_{\ell x} \end{bmatrix}
\begin{bmatrix} \llambda^{n} \\ \oomega^{n} \end{bmatrix} + \htt 
\begin{bmatrix} \rr_{\ell \lambda}^{n} + \rr_{\ell \lambda}^{n+1} \\ 0 \end{bmatrix},
\quad \forall n=1,\ldots,\nt-1,
\end{equation}
where:
\begin{equation*}
\begin{aligned}
& \llambda(t) = (\lambda_1(t), \lambda_2(t), \ldots, \lambda_m(t)), \\
& \oomega(t) = (\lambdap_1(t), \lambdap_2(t), \ldots, \lambdap_m(t)),\\
& \rr_{\ell\lambda}(t) = ({r_{\ell\lambda}}_1(t), {r_{\ell\lambda}}_2(t), \ldots, {r_{\ell\lambda}}_m(t)), 
\end{aligned}
\end{equation*}
and:
\begin{align*}
&K_{\ell x} = \left[ U_{\ell\mu_{i}}\tr K U_{\ell\mu_{j}} \right]_{1 \leqslant i, j \leqslant m}, \\
&M_{\ell x} = \left[ U_{\ell\mu_{i}}\tr M U_{\ell\mu_{j}} \right]_{1 \leqslant i, j \leqslant m}, \\
&\Mb_{\ell x} = \left[ U_{\ell\mu_{i}}\tr \Mb U_{\ell\mu_{j}} \right]_{1 \leqslant i, j \leqslant m}.
\end{align*}

\subsubsection{Hamiltonian update} 
\label{subsub:ham_update}

We consider here the system~\eqref{eq:weak-Hamilton} and solve for $(\lambda_{k}, \omega_{k})_{1 \leqslant k \leqslant m}$, with $(\mu_{k}, \nu_{k})_{1 \leqslant k \leqslant m}$ known, using the following trial and test functions:
\begin{equation*}
\begin{aligned}
&q_m(x,t) = q_{0}(x,t) + \sum_{i=1}^{m} \lambda_{i}(t) \mu_{i}(x), && \qquad
\qs(x,t) = \sum_{i=1}^{m} \lambdas_{i}(t) \mu_{i}(x), \\
&p_m(x,t) = p_{0}(x,t) + \sum_{i=1}^{m} \omega_{i}(t) \nu_{i}(x), && \qquad
\ps(x,t) = \sum_{i=1}^{m} \omegas_{i}(t) \nu_{i}(x).
\end{aligned}
\end{equation*}
Following discretization of the equations, we obtain:
\begin{equation}
\label{eq:midpoint_pgd_ham_update}
\renewcommand{\arraystretch}{1.5}
\begin{bmatrix} \htt K_{h x} & 2 C_{h x} \\ 2 C_{h x}\tr & -\htt M_{h x} \end{bmatrix} 
\begin{bmatrix} \llambda^{n+1} \\ \oomega^{n+1} \end{bmatrix} = 
\begin{bmatrix} -\htt K_{h x} & 2 C_{h x} \\ 2 C_{h x}\tr & \htt M_{h x} \end{bmatrix}
\begin{bmatrix} \llambda^{n} \\ \oomega^{n} \end{bmatrix} + \htt 
\begin{bmatrix} \rr_{h\lambda}^{n} + \rr_{h\lambda}^{n+1} \\ \rr_{h\omega}^{n} + \rr_{h\omega}^{n+1}
\end{bmatrix},
\quad \forall n=1,\ldots,\nt-1,
\end{equation}
where:
\begin{equation*}
\begin{aligned}
& \llambda(t) = (\lambda_1(t), \lambda_2(t), \ldots, \lambda_m(t)), \\
& \oomega(t) = (\omega_1(t), \omega_2(t), \ldots, \omega_m(t)),\\
& \rr_{h\omega}(t) = ({r_{h\omega}}_1(t), {r_{h\omega}}_2(t), \ldots, {r_{h\omega}}_m(t)), \\
& \rr_{h\lambda}(t) = ({r_{h\lambda}}_1(t), {r_{h\lambda}}_2(t), \ldots, {r_{h\lambda}}_m(t)), 
\end{aligned}
\end{equation*}
and:
\begin{align*}
&K_{h x} = \left[ U_{h\mu_{i}}\tr K U_{h\mu_{j}} \right]_{1 \leqslant i, j \leqslant m}, \\
&M_{h x} = \left[ U_{h\nu_{i}}\tr \Mbb U_{h\nu_{j}} \right]_{1 \leqslant i, j \leqslant m}, \\
&C_{h x} = \left[ U_{h\mu_{i}}\tr \Mb U_{h\nu_{j}} \right]_{1 \leqslant i, j \leqslant m}.
\end{align*}

\subsubsection{Gram-Schmidt process}

\clem{The question of the metric with respect to which the spatial basis should be orthogonalized in the Gram-Schmidt procedure arises: should one orthogonalize with respect to $K$, $M$, or any other  symmetric positive definite matrix? The matrices $K_{\ell x}$, $M_{\ell x}$, $\Mb_{\ell x}$, $K_{h x}$, and $M_{h x}$ introduced in the previous section have special properties.  They are called Gram matrices. Their coefficients result from scalar products with respect to discrete metrics associated with the matrices $K$, $M$, $\Mb$, or~$\Mbb$.

Let $A \in \mathbb{R}^{\nx \times \nx}$ be a symmetric positive definite matrix and let $\left( u_{1}, \ldots, u_{m} \right)$ be a family of vectors of~$\mathbb{R}^{\nx}$. One can then associate a scalar product with $A$ such that $\left<u_{i}, u_{j}\right>_{A} = u_{i}\tr A u_{j}$ and a norm such that $\| u_{i} \|_{A} = \sqrt{\left<u_{i}, u_{i}\right>_{A}}$. Let $G \in \mathbb{R}^{m \times m}$ be such that $G = \left[ \left< u_{i}, u_{j} \right>_{A} \right]_{1 \leqslant i, j \leqslant m}$, the Gram matrix associated with $A$ and $\left( u_{1}, \ldots, u_{m} \right)$. By virtue of the scalar product properties, $G$ is symmetric positive semi-definite and is invertible if and only if the vectors $\left( u_{1}, \ldots, u_{m} \right)$ are linearly independent.}

The update procedures described above strongly rely on the fact that the computed \clem{Gram} matrices are well conditioned. \clem{Yet, given the properties of the Gram matrices, some choices regarding the metric in the Gram-Schmidt procedure are more suitable than others.} In order to ensure that the condition numbers of the matrices are kept small, the Gram-Schmidt procedure is therefore performed as follows:
\begin{itemize}
\item For the Lagrangian update: \clem{one spatial basis $(U_{\ell\mu_{k}})_{1 \leqslant k \leqslant m}$ is built and orthogonalized with respect to $K$} and then normalized. In other words, \clem{following the Gram-Schmidt procedure, $K_{\ell x}$ should be equal to $I_{m}$, where $I_{m}$ is the identity matrix of size $m$. $M_{\ell x}$ and $\Mb_{\ell x}$ will not have a particular form but their conditioning numbers should remain low as long as the basis vectors remain linearly independent};
\item For the Hamiltonian update: \clem{two spatial bases $(U_{h\mu_{k}})_{1 \leqslant k \leqslant m}$ and $(U_{h\nu_{k}})_{1 \leqslant k \leqslant m}$ are built for $q$ and $p$, respectively. An optimal choice, which, to the best of our knowledge constitutes a new result, is to orthogonalize $(U_{h\mu_{k}})_{1 \leqslant k \leqslant m}$ and $(U_{h\nu_{k}})_{1 \leqslant k \leqslant m}$ with respect to $K$ and $\bar{\bar{M}}$, respectively, and then to normalize the vectors. Following the Gram-Schmidt procedure, $K_{hx} = M_{hx} = I_{m}$ and their conditioning remains optimal.}
\end{itemize}

\subsection{Adaptive fixed point algorithm}

We briefly describe in this section the fixed point algorithms for the Lagrangian and Hamiltonian formulations of the PGD approach. In particular, we propose in the case of the Hamiltonian formulation an algorithm that allows one to accelerate the convergence toward the enrichment modes associated with the generalized coordinates and the conjugate fields.

\subsubsection{Lagrangian fixed point}

For the sake of simplicity in the notation, we will simply use $\mu$ and $\lambda$ to refer here to the finite element solution $\mu_h$ (or the vector of degrees of freedom $U_{\ell \mu}$) and discrete solution $(\lambda^0,\lambda^1,\ldots,\lambda^{\nt})$. Moreover, we introduce:
\begin{itemize}
\item $\Slag : \lambda \mapsto \mu$, the operator that solves the system~\eqref{eq:disc_slag} for $\mu$ with $\lambda$ given;
\item $\Tlag : \mu \mapsto \lambda$, the operator that solves the system~\eqref{eq:midpoint_pgd_lag} for $\lambda$ with $\mu$ given.
\end{itemize}
Let $j_{\mathrm{max}}$ and $\epsilon$ denote the user-defined maximum number of iterations and tolerance. The fixed point algorithm for the Lagrangian approach is described as a pseudocode in Algorithm~\ref{algo:lag_fixed_point}. Note that the norm subscripted with $L^{2}$ is defined as follows for square-integrable functions:
\begin{equation*}
\| f \|_{L^{2}}^{2} = \myint_{\Imega}{ \myint_{\Omega}{ f(x, t)^{2}~dx } dt }
\end{equation*}

\begin{algorithm}[H]
\setstretch{1.6}
\caption{Fixed point algorithm for the Lagrangian formulation}
\label{algo:lag_fixed_point}
\begin{algorithmic}[1]
\State \textbf{Initialize} $\lambda_{0}$ and $\mu_{0}$, $j \leftarrow 0$, $s \leftarrow \epsilon + 1$
\While{$j < j_{\mathrm{max}}$ and $s > \epsilon$}
\State Increment the iteration counter: $j \leftarrow j + 1$
\State Compute new spatial mode: $\mu_{j} \leftarrow \Slag(\lambda_{j-1})$
\State Normalize: $\mu_{j} \leftarrow \mu_{j} / \| \mu_{j} \|_{K}$
\State Compute new temporal mode: $\lambda_{j} \leftarrow \Tlag(\mu_{j})$
\State Calculate the PGD difference $\Delta \leftarrow \mu_{j} \lambda_{j} - \mu_{j-1} \lambda_{j-1}$ and average $\Sigma \leftarrow \frac{1}{2} \left( \mu_{j} \lambda_{j} + \mu_{j-1} \lambda_{j-1} \right)$
\State Evaluate the stagnation coefficient: $s \leftarrow \| \Delta \|_{L^{2}} / \| \Sigma \|_{L^{2}}$
\EndWhile
\State \textbf{Return} the modes $\lambda \leftarrow \lambda_{j}$ and $\mu \leftarrow \mu_{j}$ 
\end{algorithmic}
\end{algorithm}

\subsubsection{Hamiltonian fixed point}

As before, the notations $\mu$, $\nu$, $\lambda$, $\omega$ will be used to refer to the finite element solutions $\mu_h$ and $\nu_h$ and the discrete solutions $(\lambda^0,\lambda^1,\ldots,\lambda^{\nt})$ and $(\omega^0,\omega^1,\ldots,\omega^{\nt})$.
Moreover, we consider the operators:
\begin{itemize}
\item $\Sham : (\lambda,\omega) \mapsto (\mu,\nu)$, the operator that solves the system~\eqref{eq:disc_sham} for $(\mu,\nu)$ with $(\lambda,\omega)$ given;
\item $\Tham: (\mu,\nu) \mapsto (\lambda,\omega)$, the operator that solves the system~\eqref{eq:midpoint_pgd_ham} for $(\lambda,\omega)$ with $(\mu,\nu)$ given.
\end{itemize}
We have observed in the numerical experiments that convergence of the pairs $(\mu,\lambda)$ and $(\nu,\omega)$ does not necessarily happens at the same time. We therefore propose to decouple the iterative process as follows:
\begin{itemize}
\item 
If convergence is reached on the mode $\mu\lambda$, we fix the values of $\mu$ and $\lambda$ and then solve~\eqref{eq:switchp_ham_space} and~\eqref{eq:switchp_ham_time} for $\nu$ and $\omega$ until convergence is reached:
\begin{align}
\label{eq:switchp_ham_space}
& M_{h t} \nu = C_{h t} \mu + R_{h\nu}  \\
\label{eq:switchp_ham_time}
& \htt m_{h x} \omega^{n+1} = - \htt m_{h x} \omega^{n} + 2 c_{h x} \left( \lambda^{n+1} - \lambda^{n} \right) 
+ \htt \left( r_{h\omega}^{n} + r_{h\omega}^{n+1} \right), 
\quad \forall n=0,\ldots,\nt - 1.
\end{align}
The operators associated with~\eqref{eq:switchp_ham_space} and~\eqref{eq:switchp_ham_time} will be denoted by $\Shamp : \omega \mapsto \nu$, $\Thamp : \nu \mapsto \omega$, respectively.
\item 
If convergence is reached on the mode $\nu\omega$, we fix the values of $\nu$ and $\omega$ and then solve~\eqref{eq:switchq_ham_space} and~\eqref{eq:switchq_ham_time} for $\mu$ and $\lambda$ until convergence is reached:
\begin{align}
\label{eq:switchq_ham_space} 
&K_{h t} \mu = - D_{h t} \nu + R_{h \mu} \\
\label{eq:switchq_ham_time}
&\htt k_{h x} \lambda^{n+1} 
= - \htt k_{h x} \lambda^{n} + 2 c_{h x} \left( \omega^{n} - \omega^{n+1} \right) 
+ \htt \left( r_{h \lambda}^{n} + r_{h \lambda}^{n+1} \right),
\quad \forall n=0,\ldots,\nt - 1.
\end{align}
The operators associated with~\eqref{eq:switchq_ham_space} and~\eqref{eq:switchq_ham_time} will be denoted by $\Shamq : \lambda \mapsto \mu$ and $\Thamq : \mu \mapsto \lambda$, respectively.
\end{itemize}

The fixed-point algorithm for the Hamiltonian approach is detailed as a pseudocode in Algorithm~\ref{algo:ham_fixed_point}. It is essential to notice that the dimensions of the systems to solve with the H-PGD are twice as big as the ones with the L-PGD. The advantage of this implementation is twofold. On the one hand, it allows to control the convergence of the fields separately. On the other hand, this fixed-point algorithm stops iterating on the field that has converged and therefore iterates on systems twice as small, i.e. of the same dimension as the L-PGD.

\section{Numerical results and discussion}
\label{sect:numerical_results}

\subsection{Test cases}

The objective of this section is to present several numerical examples in order to compare the PGD solutions obtained using the Lagrangian formulation and the Hamiltonian formulation. 
We shall consider in all experiments a one-dimensional bar of length $\ell=0.2$~m with material properties $E=220$~GPa, $\rho = 7000$~kg/m\textsuperscript{3}, and $A = 10^{-3}$~m\textsuperscript{2}. Unless stated otherwise, we will solve the differential equation~\eqref{eq:wave_equation} with $f(x,t) = 0$, $\forall x\in \Omega$, $\forall t \in \Imega$, with $T=1.15$~ms. Moreover, we will assume that the bar is always fixed at $x=0$, i.e.\ $u(0,t)=0$, $\forall t \in \Imega$. We shall consider five scenarios, that may differ one from the other by the choice of initial conditions $u_0(x)$ and $v_0(x)$ or the type of boundary condition at the endpoint $x=\ell$:
\begin{enumerate}
\item 
In the first case, we will consider homogeneous initial displacements and velocities, i.e.\ $u_0(x)=0$ and $v_0(x)=0$, $\forall t \in \Imega$, and the Neumann condition~\eqref{eq:NeumannBC} at $x = \ell$ where $g(t)$ is oscillating for $t \leqslant T/2$ and vanishes for $t > T/2$. In these experiments, the PGD solutions will be computed without performing the updating procedure described in Section~\ref{sub:update};
\item 
In this case, we will repeat the same experiment as above but using the updating procedure of Section~\eqref{sub:update} for the calculations of the PGD solutions;
\item 
In the third case, we keep the homogeneous initial conditions and replace the Neumann condition at $x=\ell$ by an oscillating Dirichlet condition;
\item 
This experiment will simulate the problem presented in~\eqref{eq:hom_wave_equation} for which one has the analytical solution~\eqref{eq:analytic_equation}; however, we will restrict the time interval to $T = 0.14$~ms in order to avoid the spurious oscillations that appear due to the discontinuity in the solution~\cite{hulbert,rostami};
\item 
The last case will consider the exact same scenario as in Case 2, but for the presence of an extra linear damping term in the wave equation.
\end{enumerate}

In the following, the PGD solutions obtained from the Lagrangian formalism and the Hamiltonian formalism will be referred to as ``L-PGD'' and ``H-PGD'', respectively. Moreover, we consider two versions of the Lagrangian formulation: ``L-PGD1'' uses the Crank-Nicolson scheme (also called the implicit trapezoidal rule) for time integration, as presented in the paper, while ``L-PGD2'' replaces the Crank-Nicolson scheme by the Newmark method with $\gamma = 1/2$ and $\beta = 1/4$~\cite{newmark, rixen}. The solutions in space will be approximated in terms of continuous piecewise linear polynomial functions for a total of $\nx = 224$ degrees of freedom, i.e.\ the domain $\Omega$ is decomposed into $\nx = 224$ elements of equal size. Likewise, the time interval $\Imega$ is divided into $\nt = 1025$ sub-intervals of equal size (except in Case 4 where we take $\nt = 1300$). Those values were chosen so that the discretization errors in space and in time are kept small with respect to the truncation errors from the PGD formulation.

\clem{\subsection{Comparison method and performance criteria}}

In order to assess the accuracy of the PGD solutions, we will use as reference solutions, the finite element solutions that are described in Sections~\ref{subsub:lag_fem} or~\ref{subsub:ham_fem} and obtained using the same discretization parameters $\nx = 224$ and $\nt = 1025$. Given a field $u=u(x,t)$ defined on $\Omega \times \Imega$, we denote by $\epsilon_u$ the relative error in the $L^2$ norm:
\begin{equation*}
\epsilon_u = \frac{\| u_m - u_\text{\itshape ref} \|_{L^2}}{\| u_\text{\itshape ref} \|_{L^2}}    
\end{equation*}
where $u_m$ is the PGD approximation of rank $m$ of $u$ and $u_\text{ref}$ is a very accurate reference solution.

We will study the evolution of the errors with respect to the number of modes $m$ in the PGD solutions and compare these to the errors that one obtains by  performing \textit{a posteriori} a Singular Value Decomposition on the reference solutions, except in Case 4 for which we will directly compare the PGD solutions to the analytical solution of problem~\eqref{eq:analytic_equation}. We will in particular look at the error in the energy of the bar over time as the energy (Hamiltonian) in the discrete PGD solution is supposed to remain constant when the external loading vanishes.

Furthermore, we will study the condition numbers of the Gram matrices computed during the temporal update procedure. Condition numbers of such matrices indirectly indicate how well the Gram-Schmidt procedure performs. As soon as the linear independence of the spatial basis is compromised, the procedure does not perform as well and condition numbers may significantly increase. Indeed, the vectors of the spatial basis are linearly independent if and only if the Gram matrices are invertible. More particularly, Gram matrices should be equal to the identity matrix (since the basis vectors are orthonormalized here). Thus, after the 1\textsuperscript{st} enrichment ($m = 1$), the condition numbers are equal to unity.

\newpage
\begin{algorithm}[H]
\setstretch{1.6}
\caption{Fixed point algorithm for the Hamiltonian formulation}
\label{algo:ham_fixed_point}
\begin{algorithmic}[1]
\State \textbf{Initialize} $\lambda_{0}$, $\omega_{0}$, $\mu_{0}$, and $\nu_{0}$, $j \leftarrow 0$, $s_{q} \leftarrow \epsilon + 1$, $s_{p} \leftarrow \epsilon + 1$
\While{$j < j_{\mathrm{max}}$ and $( s_{q} > \epsilon$ or $s_{p} > \epsilon)$}
\State Increment the iteration counter: $j \leftarrow j + 1$
\If{$s_{q} < \epsilon$}
\State Compute new spatial mode: $\nu_{j} \leftarrow \Shamp(\omega_{j-1})$
\State Normalize: $\nu_{j} \leftarrow \nu_{j} / \|\nu_{j}\|_{M}$
\State Compute new temporal mode: $\omega_{j} \leftarrow \Thamp(\nu_{j})$
\State Calculate the PGD difference $\Delta_{p} \leftarrow \nu_{j} \omega_{j} - \nu_{j-1} \omega_{j-1}$
\State and average $\Sigma_{p} \leftarrow \frac{1}{2} \left( \nu_{j} \omega_{j} + \nu_{j-1} \omega_{j-1} \right)$
\State Evaluate the stagnation coefficient: $s_{p} \leftarrow \| \Delta_{p} \|_{L^{2}} / \| \Sigma_{p} \|_{L^{2}}$
\State Update $\mu_j \leftarrow \mu_{j-1}$ and $\lambda_j \leftarrow \lambda_{j-1}$ (fixed modes)

\ElsIf{$s_{p} < \epsilon$}
\State Compute new spatial mode: $\mu_{j} \leftarrow \Shamq(\lambda_{j-1})$
\State Normalize: $\mu_{j} \leftarrow \mu_{j} / \|\mu_{j}\|_{K}$
\State Compute new temporal mode: $\lambda_{j} \leftarrow \Thamq(\mu_{j})$
\State Calculate the PGD difference $\Delta_{q} \leftarrow \mu_{j} \lambda_{j} - \mu_{j-1} \lambda_{j-1}$
\State and average $\Sigma_{q} \leftarrow \frac{1}{2} \left( \mu_{j} \lambda_{j} + \mu_{j-1} \lambda_{j-1} \right)$
\State Evaluate the stagnation coefficient: $s_{q} \leftarrow \| \Delta_{q} \|_{L^{2}} / \| \Sigma_{q} \|_{L^{2}}$
\State Update $\nu_j \leftarrow \nu_{j-1}$ and $\omega_j \leftarrow \omega_{j-1}$ (fixed modes)
\Else
\State Calculate new spatial modes: $(\mu_{j}, \nu_{j}) \leftarrow \Sham(\lambda_{j-1}, \omega_{j-1})$
\State Normalize: $\mu_{j} \leftarrow \mu_{j} / \|\mu_{j}\|_{K}$ and $\nu_{j} \leftarrow \nu_{j} / \|\nu_{j}\|_{M}$
\State Compute new temporal modes: $(\lambda_{j}, \omega_{j}) \leftarrow \Tham(\mu_{j}, \nu_{j})$
\State Calculate the PGD differences $\Delta_{q} \leftarrow \mu_{j} \lambda_{j} - \mu_{j-1} \lambda_{j-1}$ and $\Delta_{p} \leftarrow \nu_{j} \omega_{j} -  \nu_{j-1} \omega_{j-1}$
\State and averages $\Sigma_{q} \leftarrow \frac{1}{2} \left( \mu_{j} \lambda_{j} + \mu_{j-1} \lambda_{j-1} \right)$ and $\Sigma_{p} \leftarrow \frac{1}{2} \left( \nu_{j} \omega_{j} + \nu_{j-1} \omega_{j-1} \right)$
\State Evaluate the stagnation coefficient: $s_{q} \leftarrow \| \Delta_{q} \|_{L^{2}} / \| \Sigma_{q} \|_{L^{2}}$ and $s_{p} \leftarrow \| \Delta_{p} \|_{L^{2}} / \| \Sigma_{p} \|_{L^{2}}$
\EndIf
\EndWhile
\State \textbf{Return} the modes $\lambda = \lambda_{j}$, $\omega = \omega_{j}$, $\mu = \mu_{j}$, and $\nu = \nu_{j}$
\end{algorithmic}
\end{algorithm}

Then, when additional enrichments are considered, two scenarios may occur:
\begin{enumerate}
\item 
The Gram-Schmidt procedure performs well, the Gram matrices remain equal to $I_{m}$ (the identity matrix of size $m$), and the condition numbers remain equal to unity;
\item 
Linear independence is compromised and not only some of the off-diagonal coefficients of the Gram matrices become non-null but the Gram matrices are no longer invertible. As a result the condition numbers drastically increase.
\end{enumerate}

\begin{figure}[tb]
    \centering
    \includegraphics[width=\textwidth]{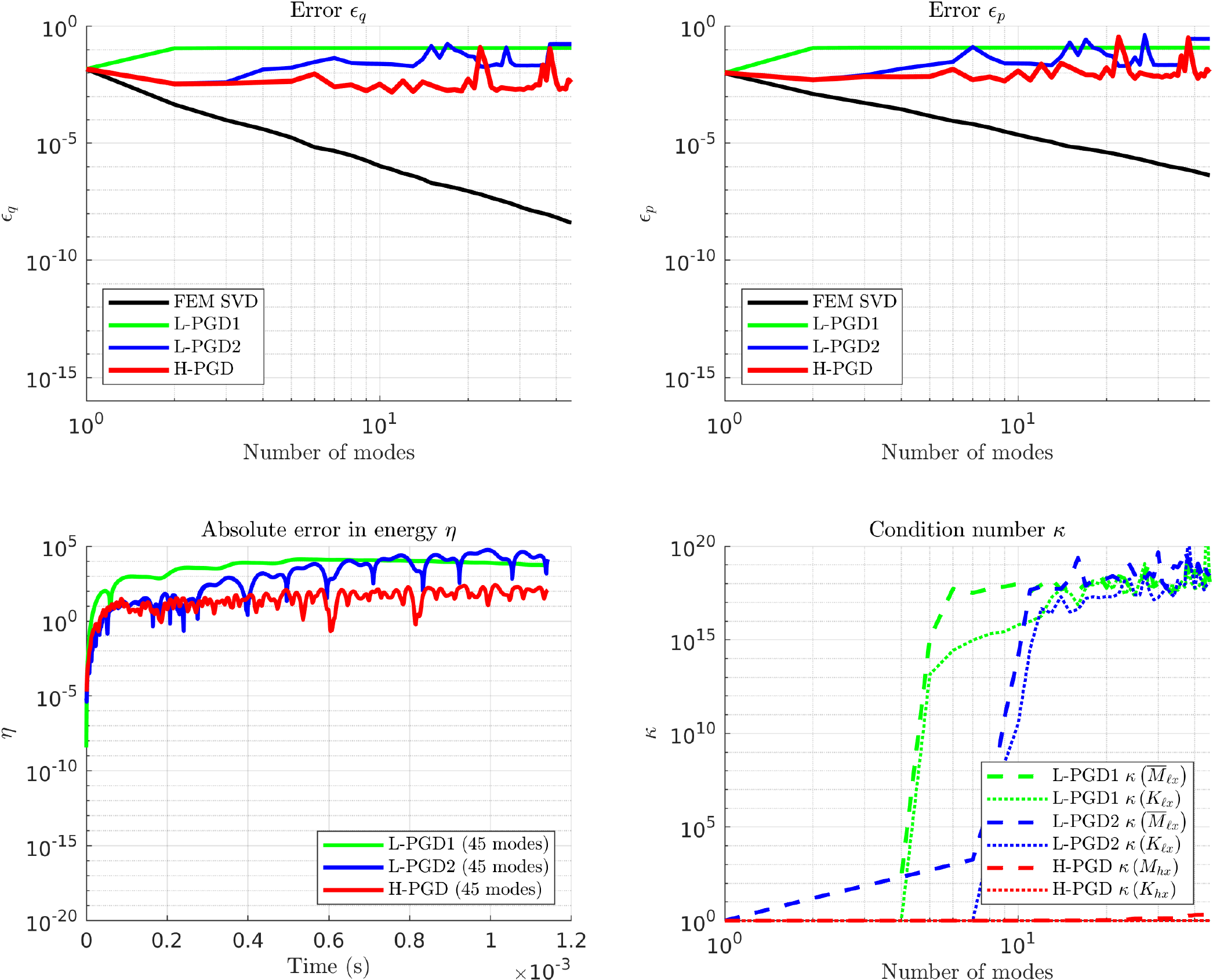}
    \caption{Case 1. (Top left) Error between the reference displacement field and the SVD or PGD displacement field. (Top right) Error between the reference conjugate momenta field and the SVD or PGD conjugate momenta field. (Bottom left) Error between the energy of the reference system and the energy of the reduced system over time. (Bottom left) Condition numbers of the matrices introduced in Sections~\ref{subsub:lag_update} and~\ref{subsub:ham_update}.}
    \label{fig:dir-neum-noup-plot}
\end{figure}

\subsection{Case 1: Neumann BC without updating procedure}

We approximate in this case Problem~\eqref{eq:wave_equation}-\eqref{eq:NeumannBC} with $u_0=0$, $v_0=0$, and the Neumann boundary condition:
\begin{equation*}
EA \frac{\partial u}{\partial x} (\ell,t) = g(t) 
= \left\{ 
\begin{aligned}
&F_{0} \left( 1 - \cos \left( \omega t \right) \right), 
&&\quad \forall t \in (0,T/2], \\
&0, 
&&\quad \forall t \in (T/2,T). 
\end{aligned} 
\right.
\end{equation*}
where $F_0=10^{6}$~N and $\omega = 4.4 \times 10^{4}$~rad/s.

We observe in Figure~\ref{fig:dir-neum-noup-plot} that the $L^{2}$ errors in the generalized coordinates and momenta of the PGD solutions barely decrease, if at all, and that their evolution is non monotonic. However, the H-PGD solution seems to behave slightly better than the L-PGD solutions, especially in terms of the absolute error in energy that remains smaller. We also observe that the condition numbers $\kappa(K_{\ell x})$ and $\kappa(\Mb_{\ell x})$ associated with the matrices $K_{\ell x}$ and $\Mb_{\ell x}$ increase as soon as the 5\textsuperscript{th} for L-PGD1 and as soon as the 8\textsuperscript{th} enrichment for L-PGD2. This indicates that the Gram-Schmidt algorithm fails to orthogonalize the spatial basis. This is not due to numerical instability and it therefore cannot be corrected by the modified Gram-Schmidt algorithm. The increase in the condition number is a consequence of the degeneration of the basis associated with the spatial modes: the added modes compromise the linear independence of the modes. In other words, the new modes do not provide any new information that was not already contained in the previous decomposition.

\begin{figure}[tb]
    \centering
    \includegraphics[width=\textwidth]{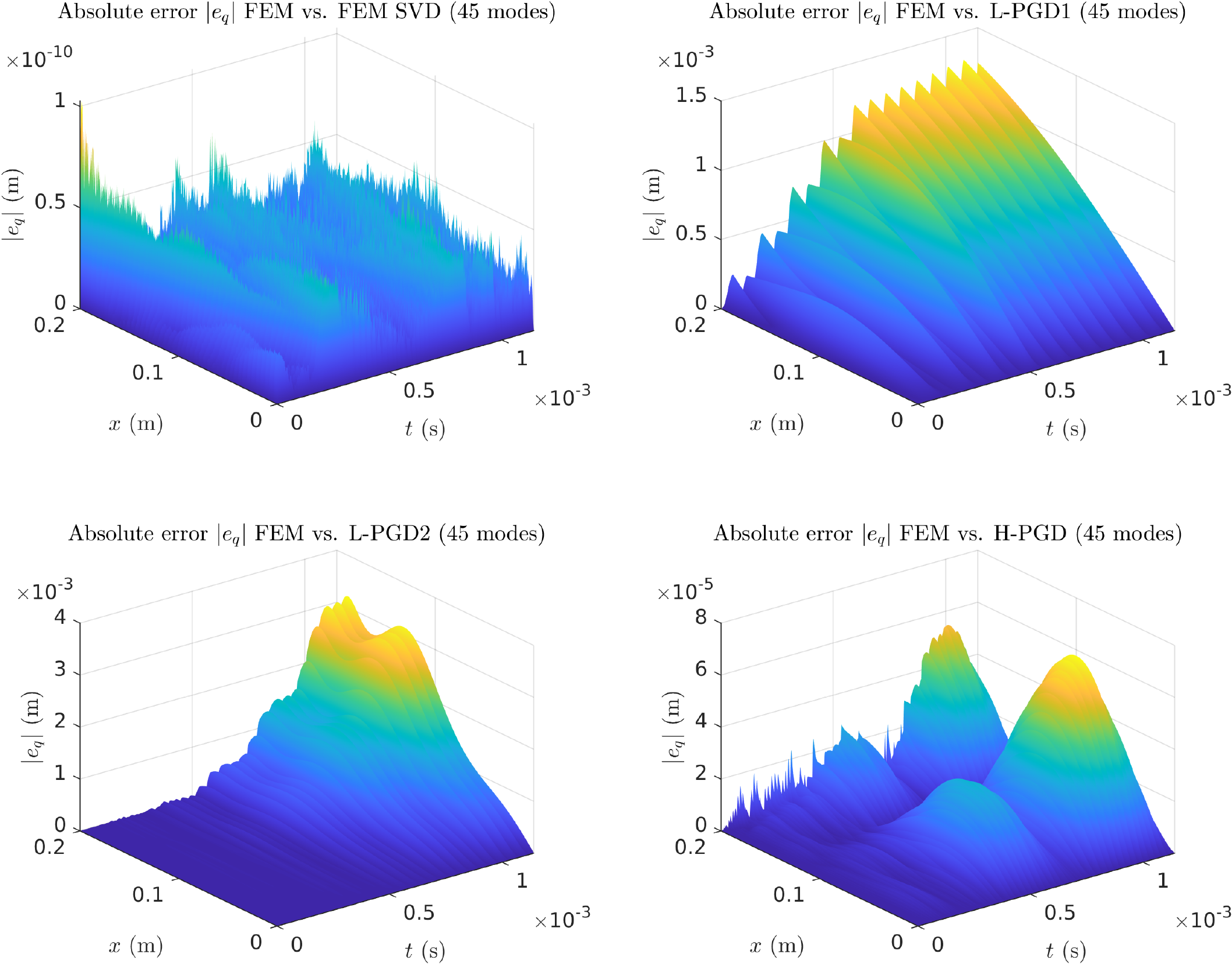}
    \caption{Case 1. Absolute errors in space and time between the reference displacement field and the SVD or PGD displacement field.}
    \label{fig:dir-neum-noup-q}
\end{figure}

\begin{figure}[H]
    \centering
    \vspace{.2in}
    \includegraphics[width=\textwidth]{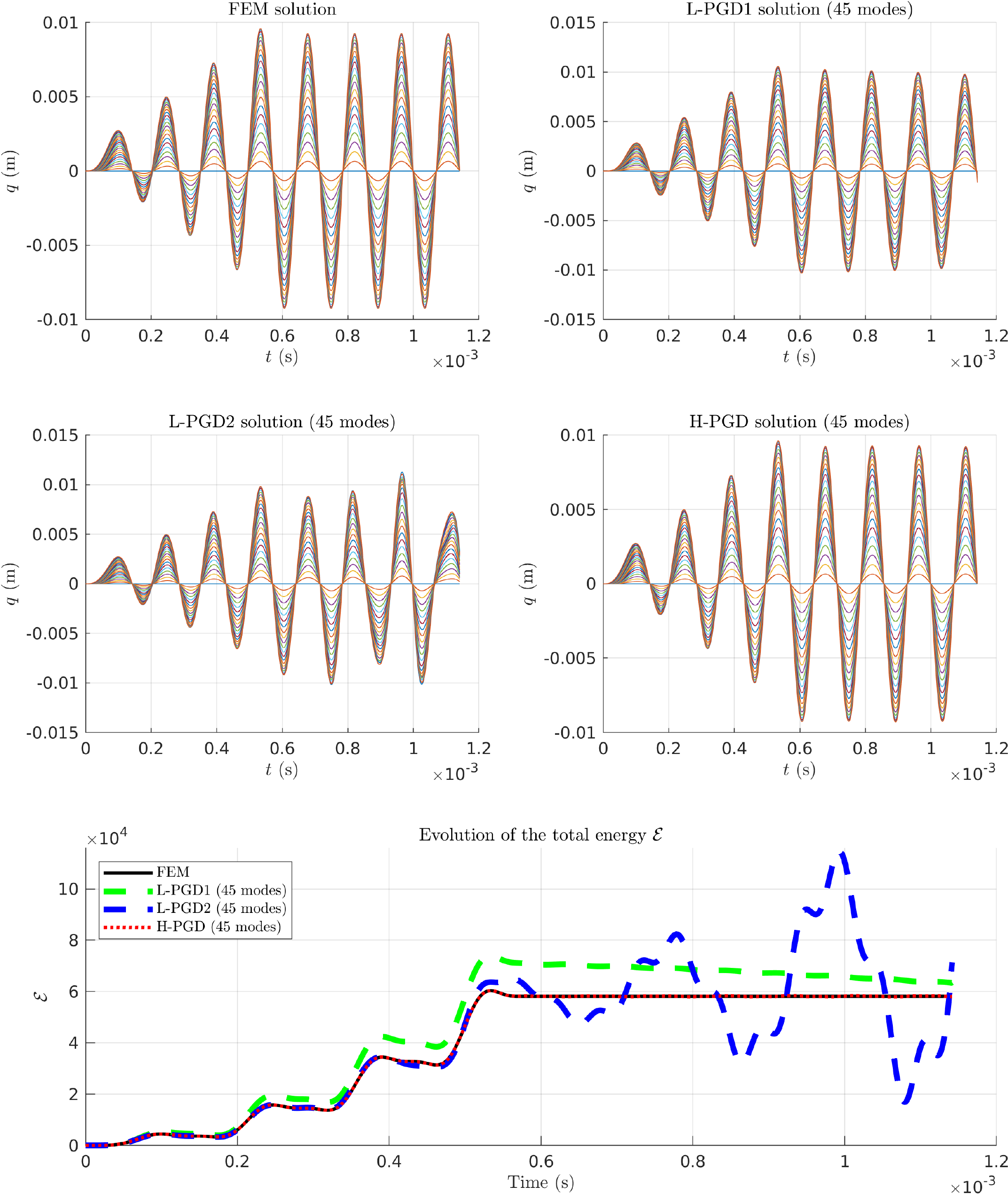}
    \caption{Case 1. (Top four plots) Evolution of the displacement field over time for the different reduction methods; the displacements are shown at 23 nodes uniformly distributed along the bar. (Bottom) Evolution of the total  energy of the reference and reduced systems versus time.}
    \label{fig:dir-neum-noup-energy}
\end{figure}

Figure~\ref{fig:dir-neum-noup-q} shows the distribution of the errors in space and time for the PGD solutions while Figure~\ref{fig:dir-neum-noup-energy} illustrate the evolution of the displacement fields over time and of the total energy for the FEM solution and the different PGD solutions.

\begin{figure}[tb]
    \centering
    \includegraphics[width=\textwidth]{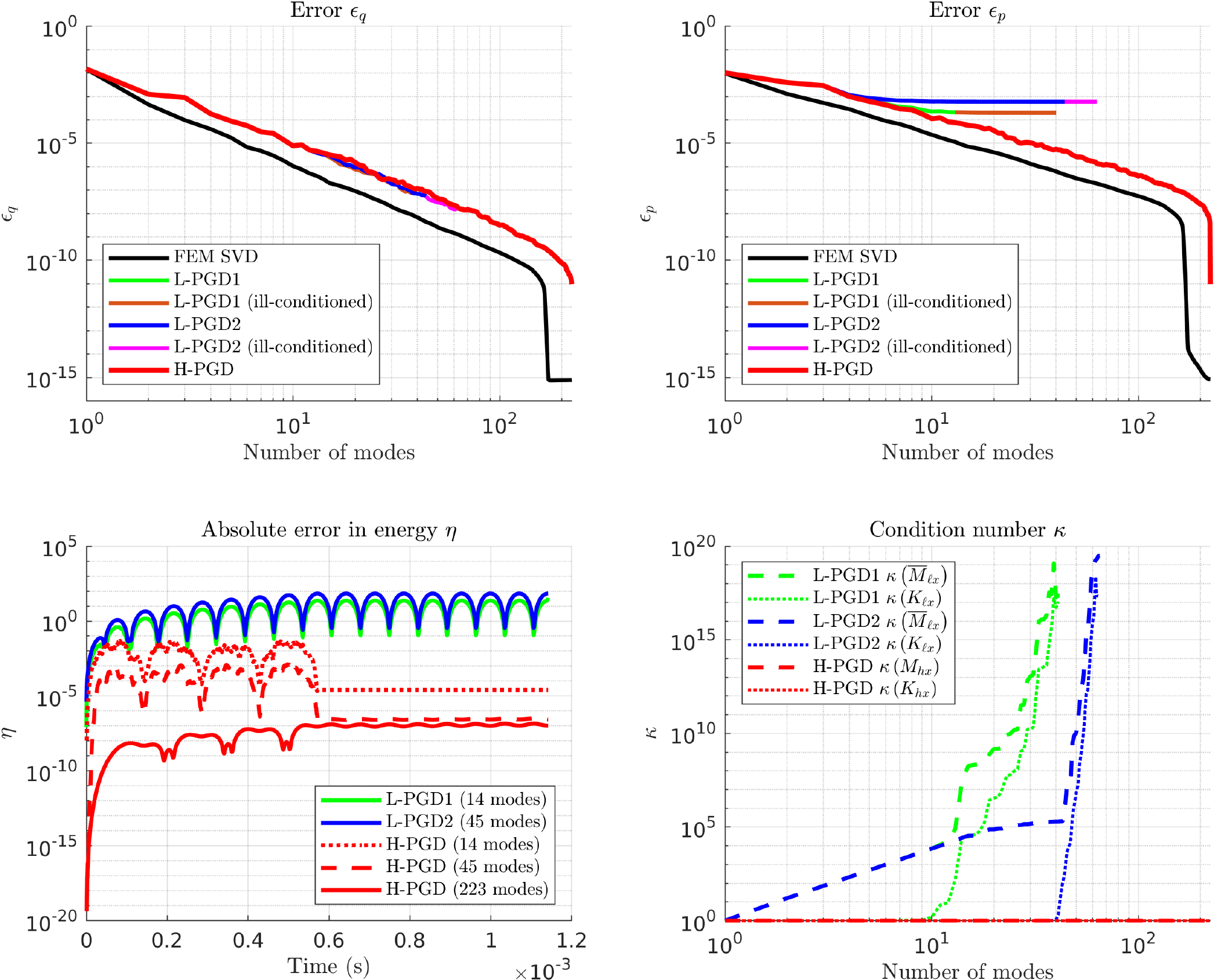}
    \caption{Case 2. (Top left) Error between the reference displacement field and the SVD or PGD displacement field. (Top right) Error between the reference conjugate momenta field and the SVD or PGD conjugate momenta field. (Bottom left) Error between the energy of the reference system and the energy of the reduced system over time. (Bottom left) Condition numbers of the matrices introduced in Sections~\ref{subsub:lag_update} and~\ref{subsub:ham_update}.} 
    \label{fig:dir-neum-plot}
\end{figure}

\begin{figure}[tb]
    \centering
    \includegraphics[width=\textwidth]{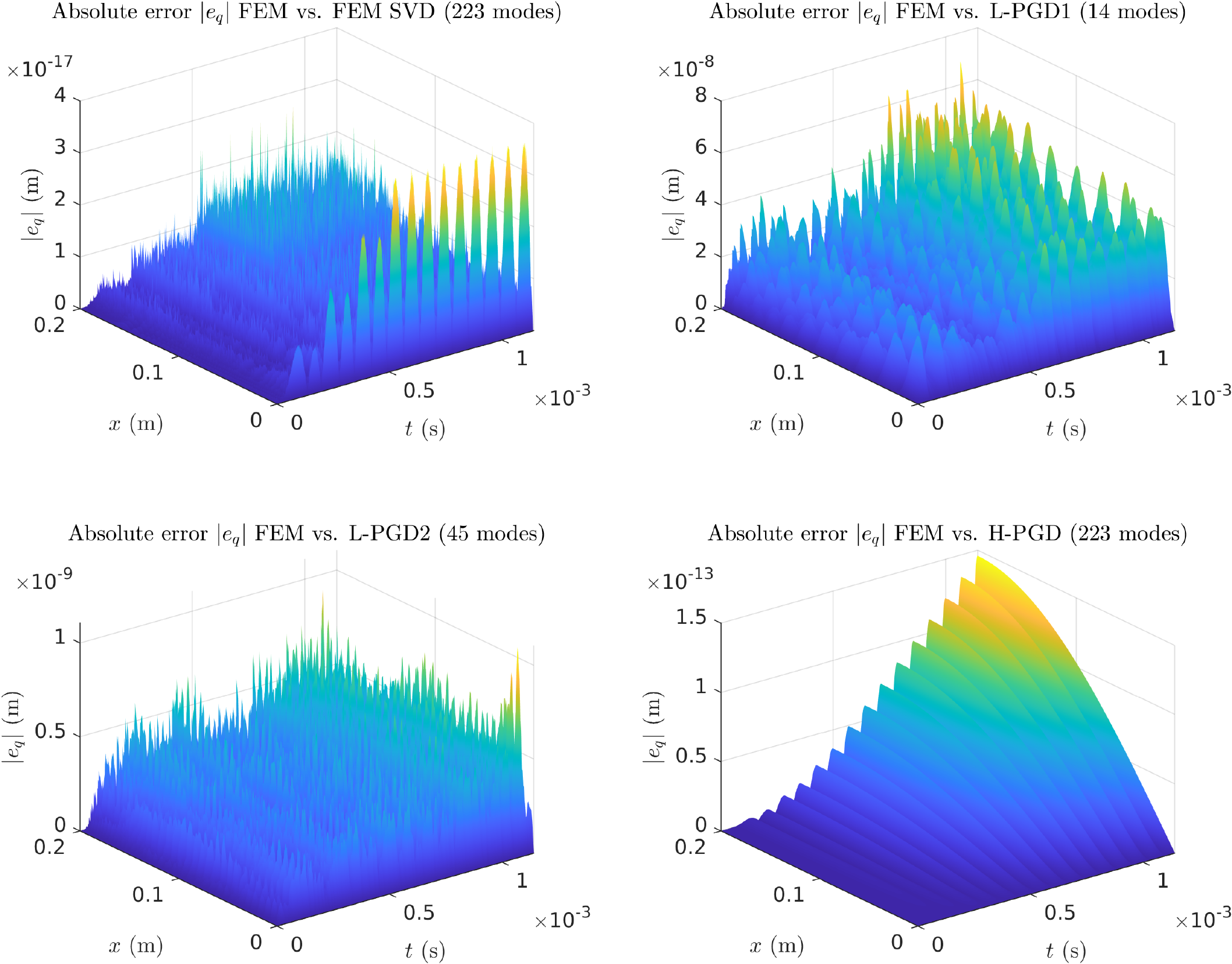}
    \caption{Case 2. Absolute errors in space and time between the reference displacement field and the SVD or PGD displacement field.} 
    \label{fig:dir-neum-q}
\end{figure}

\subsection{Case 2: Neumann BC with updating procedure} 
\label{subsub:dir-neu}

We repeat here the same experiment of Case 1 using this time the updating procedure of Section~\eqref{sub:update} for the calculations of the PGD solutions. We show in Figure~\ref{fig:dir-neum-plot} the errors in $L^2$ norm and energy and the condition numbers of the matrices introduced in Sections~\ref{subsub:lag_update} and~\ref{subsub:ham_update}. We first point out that the updating procedure significantly improves the convergence. Nevertheless, we observe that in the case of the Lagrangian PGD solutions, the matrices for L-PGD1 and L-PGD2 become ill-conditioned as soon as the 14 and 45 modes are reached, respectively. For the L-PGD1 and L-PGD2, the space modes $\mu_{k}$ are orthogonalized and normalized with respect to Matrix $K$. Thus, the condition number of $K_{\ell x}$ remains low for a dozen of modes (as long as $K_{\ell x} = I_{m}$) while the condition number of $\Mb_{\ell x}$ increases from the beginning. It follows that the condition numbers diverge for the L-PGD1 and L-PGD2 around 40 and 60 modes, respectively.
 
On the other hand, it is remarkable that the matrices in the case of the H-PGD solution always remain well-conditioned. This is explained by the fact that at each enrichment step, H-PGD manages to compute a new mode whose information is not already contained in the old spatial modes. In other words, the Gram-Schmidt algorithm manages to enforce $M_{h x} = I_{m}$ and $K_{h x} = I_{m}$.

Before divergence of the L-PGD solutions occurs, we observe that the errors $\epsilon_{q}$ in the $L^2$ norm in the displacement field follow the same behavior for the three PGDs. However, for L-PGD, the errors in the conjugate momenta $\epsilon_{p}$ quickly reach a plateau after the calculation of the first dozen modes due to the fact that the matrices become ill-conditioned. It follows that the L-PGD approach fails to identify the relevant modes for $p$. For the H-PGD approach, we see that $\epsilon_{p}$ keeps decreasing since the method is explicitly designed to compute separate decompositions for both $q$ and $p$. It also implies that the total energy of the system is well approximated for the $223$ modes of the H-PGD solution unlike in the case of the L-PGD solutions. We actually observe in Figure~\ref{fig:dir-neum-q} that the distribution of the errors in space and time for the H-PGD solution remain a few orders of magnitude lower than for the L-PGD solutions, even after the calculation of the $223$ modes.

\begin{figure}[H]
    \centering
    \vspace{.2in}
    \includegraphics[width=\textwidth]{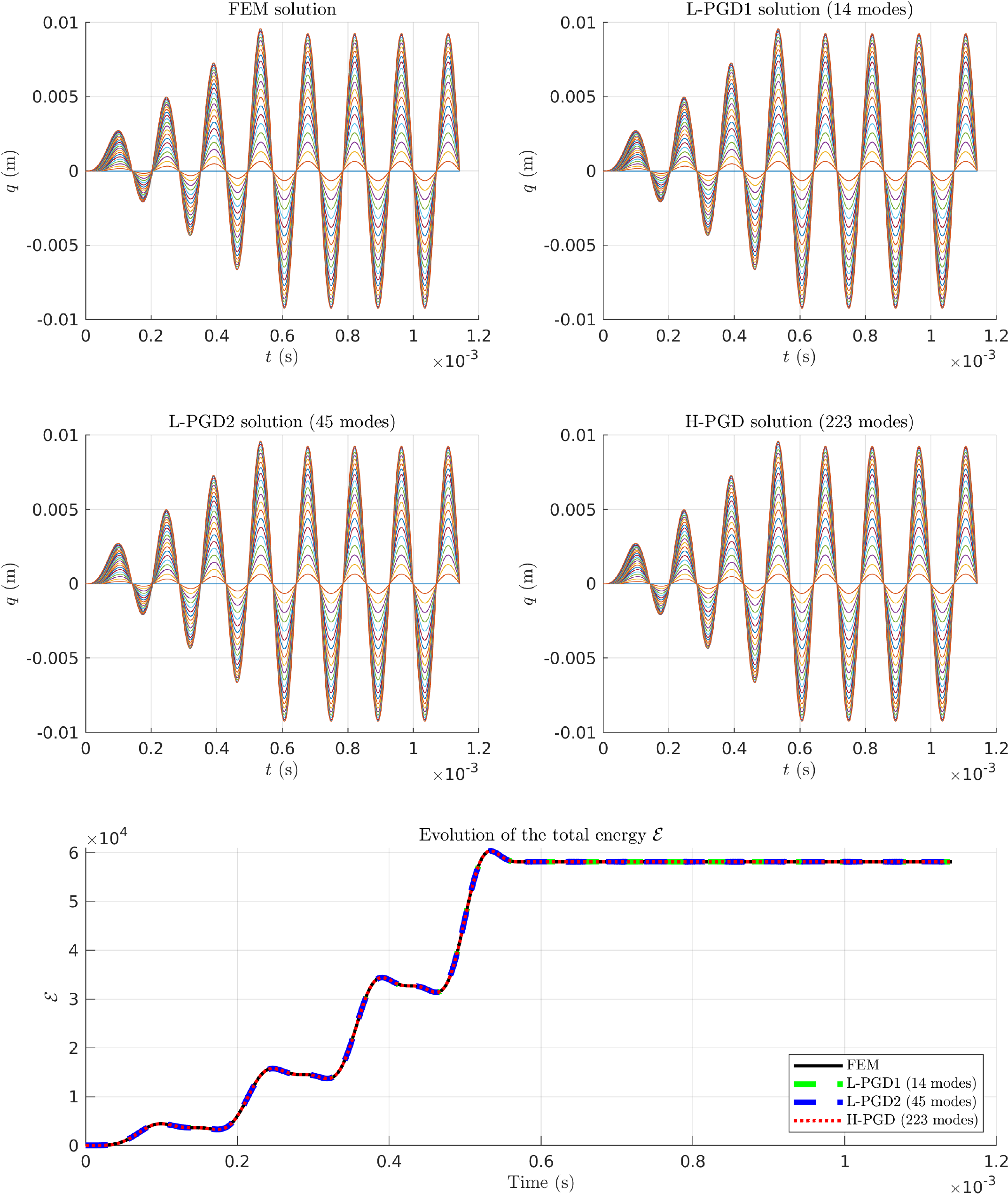}
    \caption{Case 2. (Top four plots) Evolution of the displacement field over time for the different reduction methods; the displacements are shown at 23 nodes uniformly distributed along the bar. (Bottom) Evolution of the total  energy of the reference and reduced systems versus time.} 
    \label{fig:dir-neum-energy}
\end{figure}

Finally, we show in Figure~\ref{fig:dir-neum-energy} the evolution of the displacement fields over time and of the total energy for the FEM solution and the different PGD solutions. We note that we use in these plots only the first $14$ modes for L-PGD1, $45$ modes for L-PGD2, and the total of $223$ modes for H-PGD. We see that the energy of the system increases as long as the force applied at the end of the beam is non-zero and remains constant once the end of the beam becomes free, as expected. In other words, the Hamiltonian of the system, i.e.\ the total energy is preserved when the system is conservative.

\begin{figure}[tb]
    \centering
    \includegraphics[width=\textwidth]{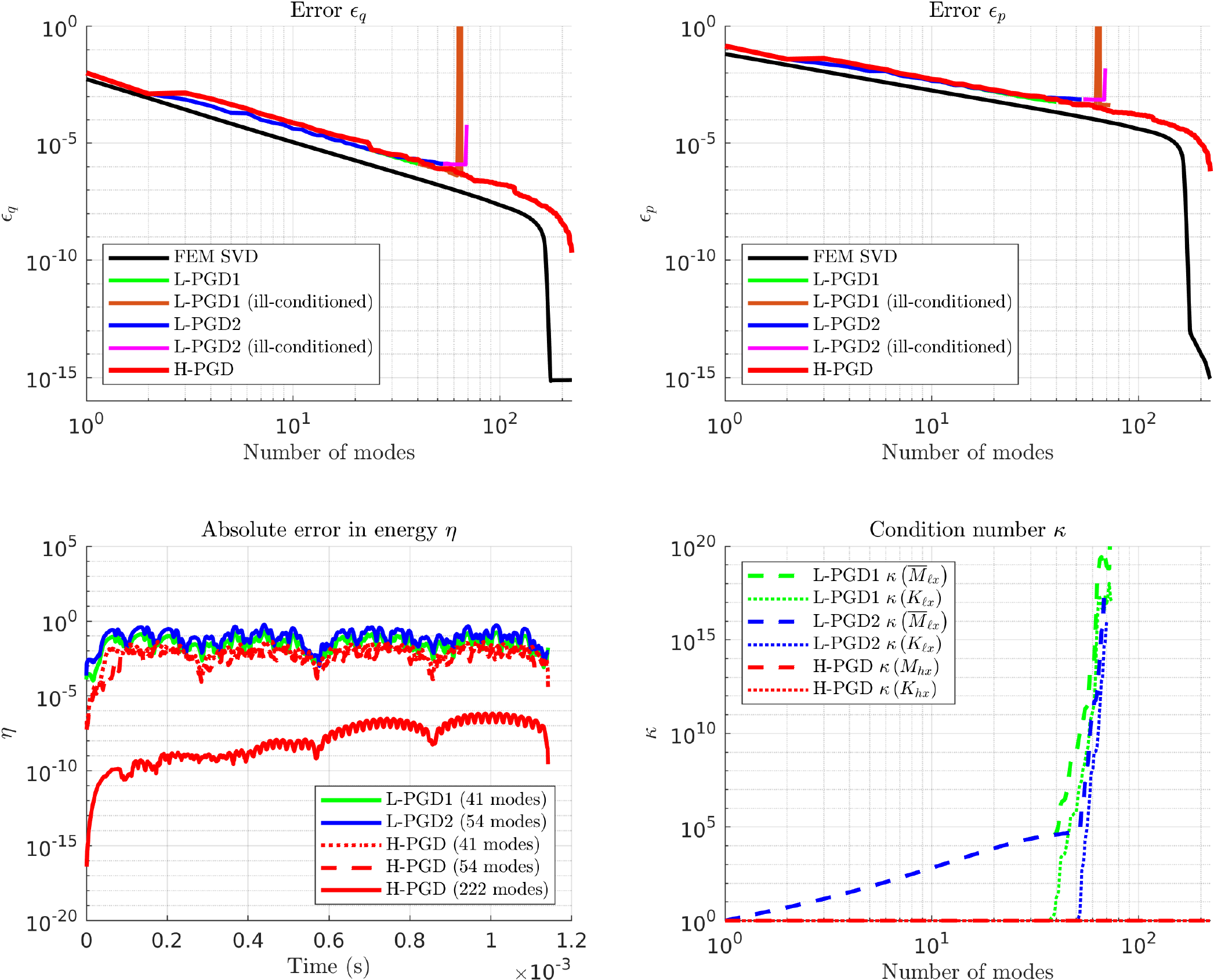}
    \caption{Case 3. (Top left) Error between the reference displacement field and the SVD or PGD displacement field. (Top right) Error between the reference conjugate momenta field and the SVD or PGD conjugate momenta field. (Bottom left) Error between the energy of the reference system and the energy of the reduced system over time. (Bottom left) Condition numbers of the matrices introduced in Sections~\ref{subsub:lag_update} and~\ref{subsub:ham_update}.} 
    \label{fig:dir-dir-plot}
\end{figure}

\subsection{Case 3: Oscillating Dirichlet BC}

In this section, we replace the Neumann boundary condition at the end point $x=\ell$ in the previous problem by the oscillatory Dirichlet boundary condition:
\begin{equation*}
u(\ell, t) = U_{0} \left( 1 - \cos \left( \omega t \right) \right), 
\qquad \forall t \in \Imega,
\end{equation*}
where $U_0 = 5$~mm and $\omega = 1.1 \times 10^{4}$~rad/s.

\begin{figure}[tb]
    \centering
    \includegraphics[width=\textwidth]{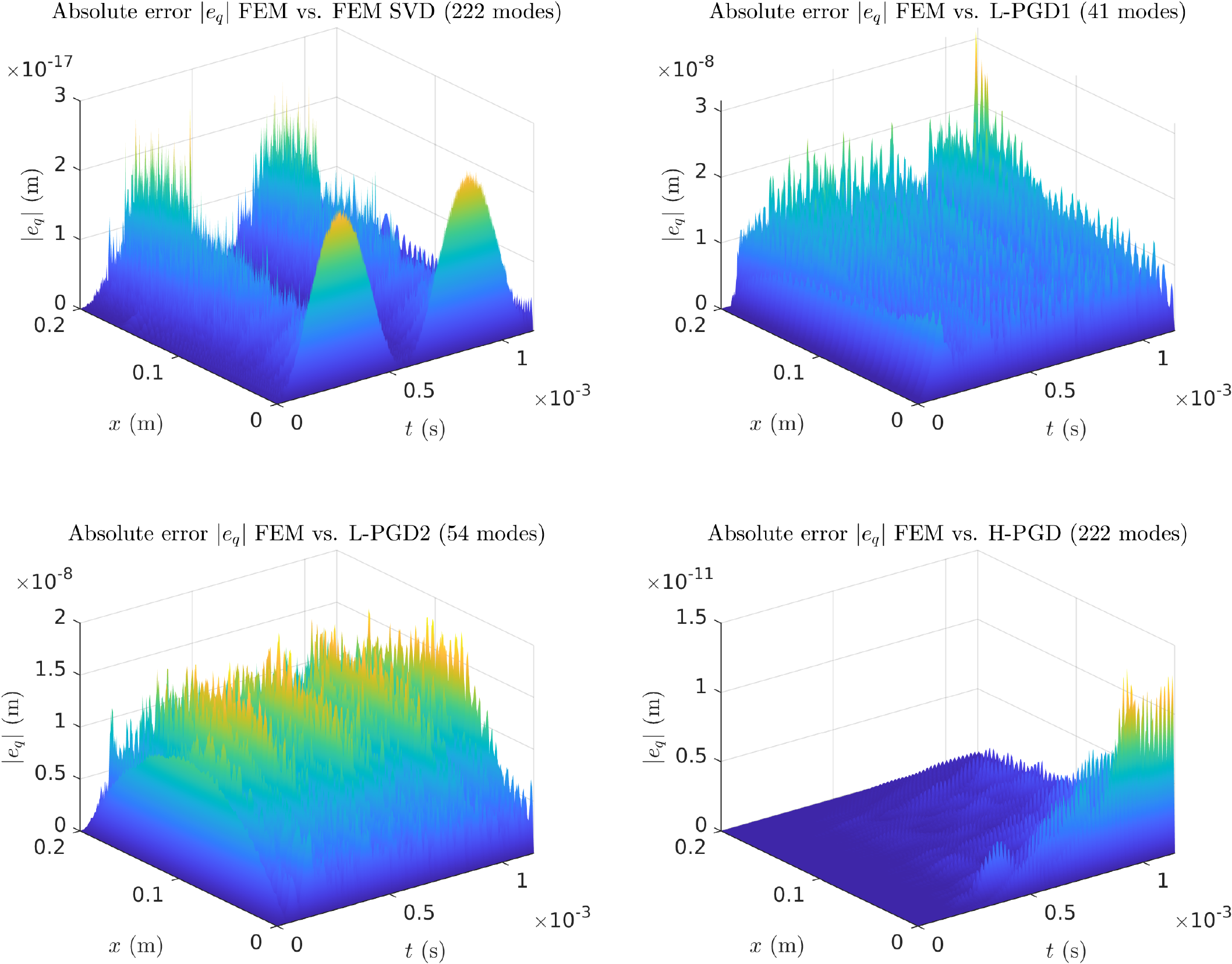}
    \caption{Case 3. Absolute errors in space and time between the reference displacement field and the SVD or PGD displacement field.} 
    \label{fig:dir-dir-q}
\end{figure}

\begin{figure}[H]
    \centering
    \vspace{.2in}
    \includegraphics[width=\textwidth]{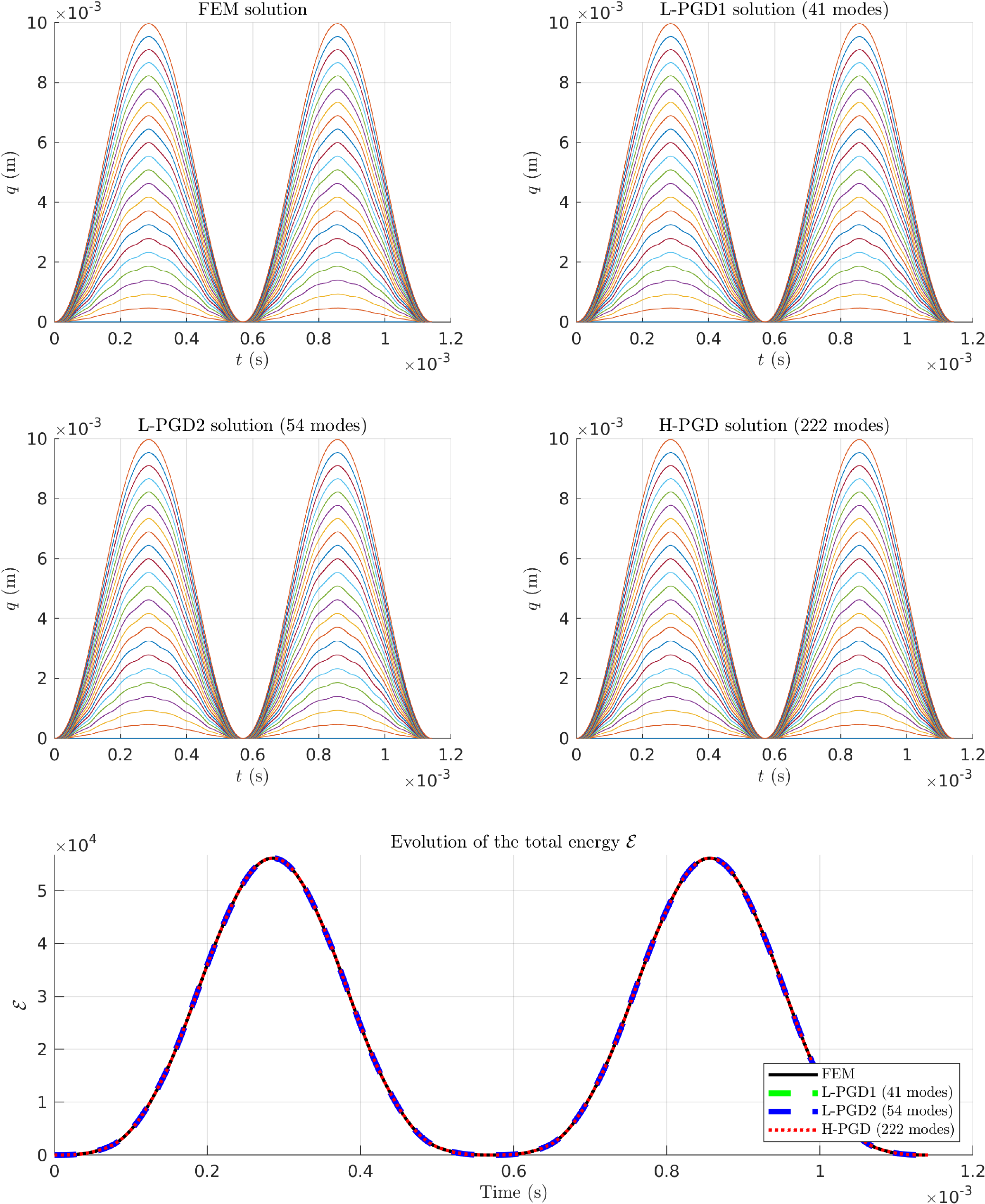}
    \caption{Case 3. (Top four plots) Evolution of the displacement field over time for the different reduction methods; the displacements are shown at 23 nodes uniformly distributed along the bar. (Bottom) Evolution of the total energy of the reference and reduced systems versus time.} 
    \label{fig:dir-dir-energy}
\end{figure}

We collect the numerical results in  Figures~\ref{fig:dir-dir-plot},~\ref{fig:dir-dir-q} and~\ref{fig:dir-dir-energy}. We essentially observe the same behaviors as in the previous test case, except that the matrices associated with the Lagrangian approaches become ill-conditioned after a larger number of computed modes than before and that the relative errors $\epsilon_q$ in the displacement and $\epsilon_{p}$ in the conjugate momenta quickly diverge rather than reaching a plateau. It is also clear from these results that the H-PGD formulation produces superior results in terms of convergence and accuracy.

\subsection{Case 4: Comparison with analytical solution}

\begin{figure}[H]
    \centering
    \vspace{.2in}
    \includegraphics[width=\textwidth]{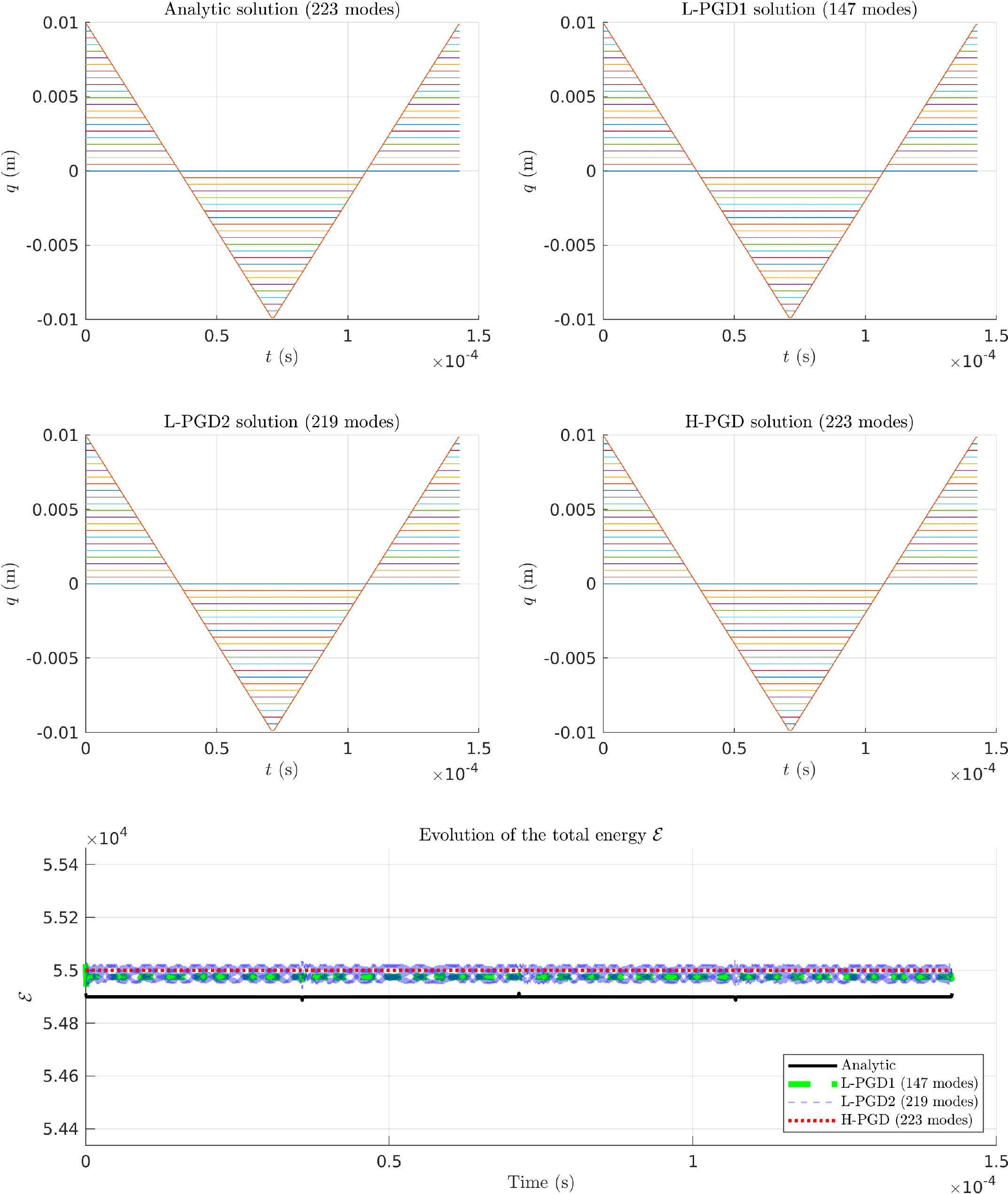}
    \caption{Case 4. (Top four plots) Evolution of the displacement field over time for the different reduction methods; the displacements are shown at 23 nodes uniformly distributed along the bar. (Bottom) Evolution of the total  energy of the reference and reduced systems versus time.}
    \label{fig:shock-energy}
\end{figure}

We solve in this section the problem formulated in~\eqref{eq:hom_wave_equation} where the initial displacement is given by $u_0(x) = Fx/(EA)$, $\forall x \in \Omega$, with $F/(EA) = 0.05$. This test case describes a shock-type wave featuring a discontinuity in the first derivative, see Figure~\ref{fig:shock-energy}.

\begin{figure}[tb]
    \centering
    \includegraphics[width=\textwidth]{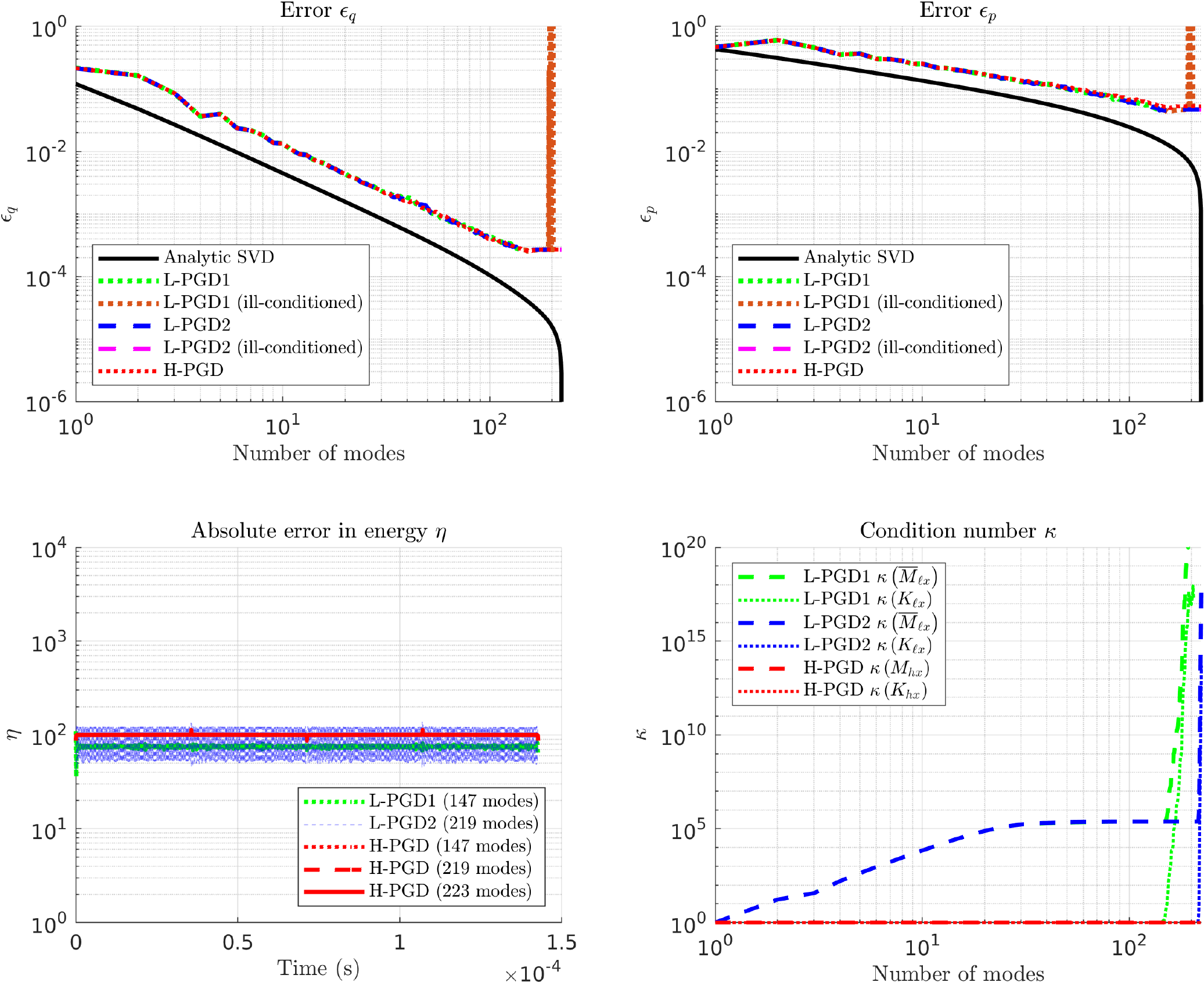}
    \caption{Case 4. (Top left) Error between the reference displacement field and the SVD or PGD displacement field. (Top right) Error between the reference conjugate momenta field and the SVD or PGD conjugate momenta field. (Bottom left) Error between the energy of the reference system and the energy of the reduced system over time. (Bottom left) Condition numbers of the matrices introduced in Sections~\ref{subsub:lag_update} and~\ref{subsub:ham_update}.}
    \label{fig:shock-plot}
\end{figure}

Errors and condition numbers for this test case are shown in Figure~\ref{fig:shock-plot}. We observe that the matrices for the L-PGD approaches eventually become ill-conditioned again. Nevertheless, the errors for the three PGD seem to decrease at the same rate. In the analytical solution provided in~\eqref{eq:analytic_equation}, the error in the truncated displacement retaining only the first $m$ modes is of the order $\mathcal{O}(1/m^{2})$. In comparison, the error in the computed PGDs seems to be approximately of the order $\mathcal{O}(1/m^{\frac{3}{2}})$.

\begin{figure}[tb]
    \centering
    \includegraphics[width=\textwidth]{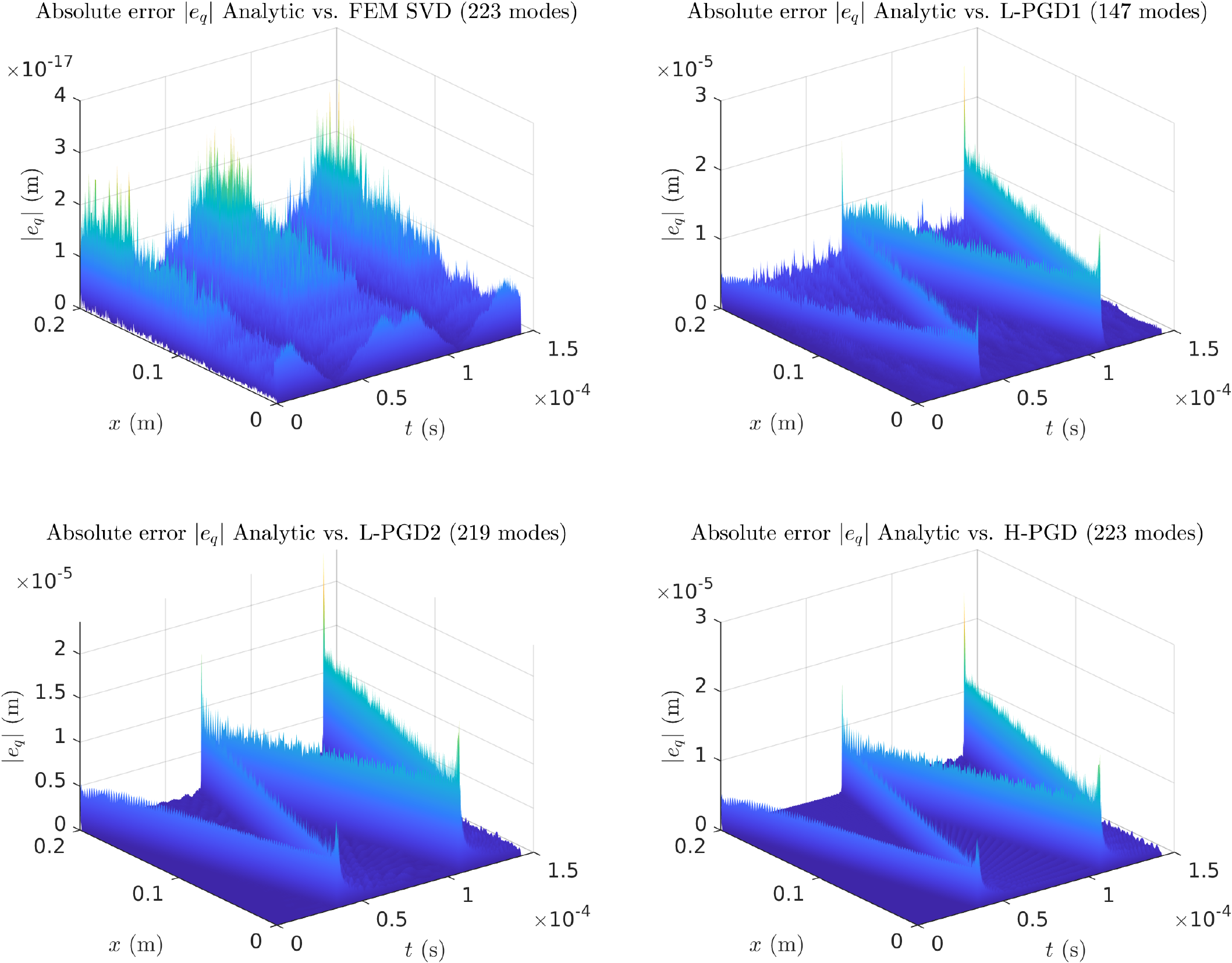}
    \caption{Case 4. Absolute errors in space and time between the reference displacement field and the SVD or PGD displacement field.}
    \label{fig:shock-q}
\end{figure}

We show in Figure~\ref{fig:shock-q} the distribution of the absolute errors in space and time for the FEM SVD solution, the L-PGD1 and L-PGD2 solutions, and the H-PGD solution. We observe on the one hand that the errors in the FEM SVD solution are negligible. On the other hand, the errors in the last three solutions are of the same order and locally concentrated around the wavefront.

These errors are essentially due to the presence of the discontinuity in the solution, which make them difficult to capture. This is out of the scope of this study but we mention that the use of a time-discontinuous Galerkin (TGD) integration scheme could possibly address this issue~\cite{boucinha2013,hulbert}.
As a last remark, it seems that the H-PGD solution away from the location of the discontinuities seems to be less polluted by the large errors than for the L-PGD solutions, in particular in the vicinity of the end point $x=\ell$.

\subsection{Case 5: Damped bar with Neumann BC}

The last case considers the exact same scenario presented in Section~\ref{subsub:dir-neu}, but for the presence of an extra linear damping term in the wave equation, i.e.
\begin{equation*}
\rho A \frac{\partial^{2} u}{\partial t^{2}} + \zeta \frac{\partial u}{\partial t} - EA \frac{\partial^{2} u}{\partial x^{2}} = 0, \qquad \forall (x,t) \in \Omega \times \Imega,
\end{equation*}
where $\zeta = 15 \times 10^{3}$~Pl (1 Poiseuille = 1 kg/m/s) is the so-called damping coefficient.

\begin{figure}[tb]
    \centering
    \includegraphics[width=\textwidth]{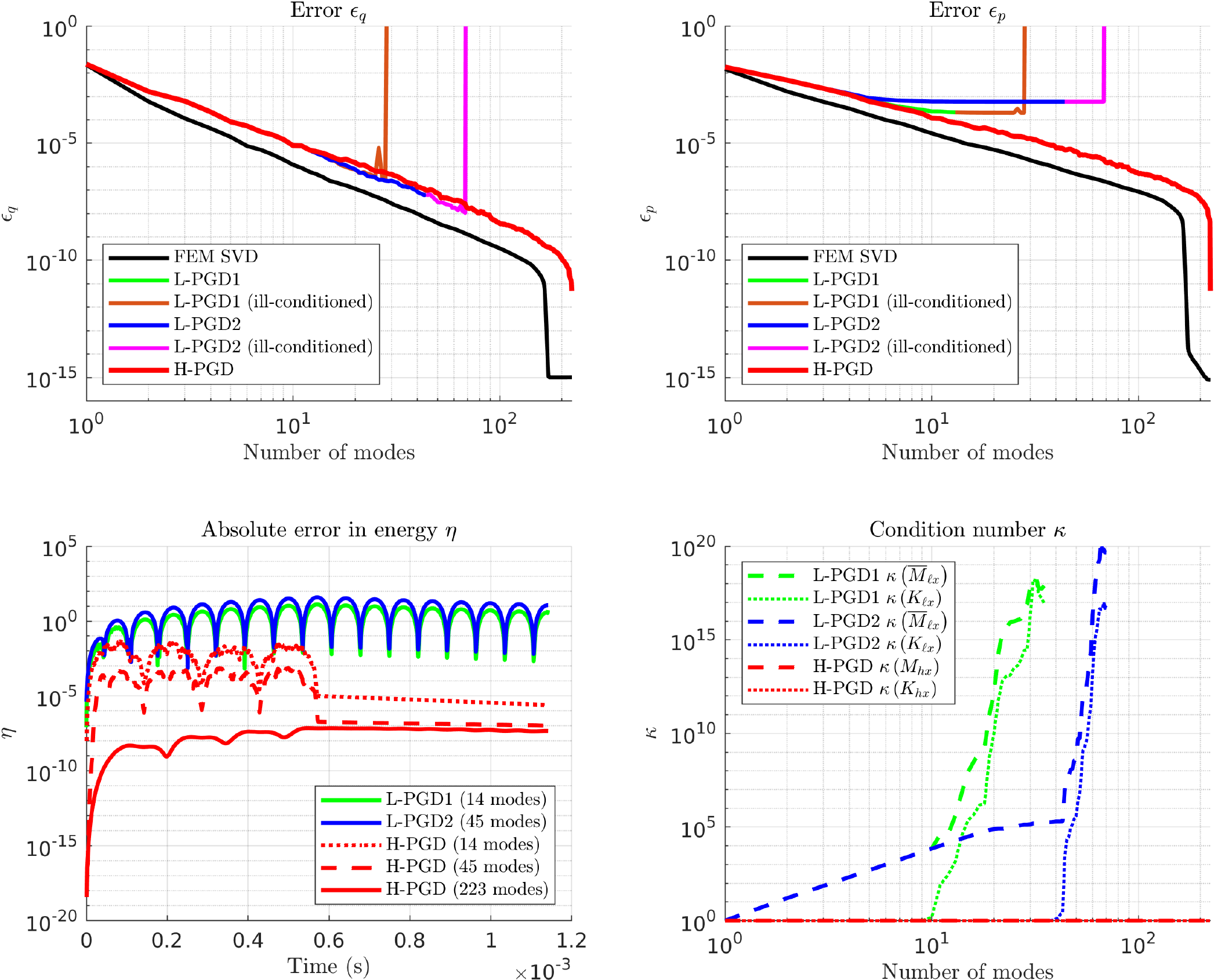}
    \caption{Case 5. (Top left) Error between the reference displacement field and the SVD or PGD displacement field. (Top right) Error between the reference conjugate momenta field and the SVD or PGD conjugate momenta field. (Bottom left) Error between the energy of the reference system and the energy of the reduced system over time. (Bottom left) Condition numbers of the matrices introduced in Sections~\ref{subsub:lag_update} and~\ref{subsub:ham_update}.} 
    \label{fig:damp-dir-neum-plot}
\end{figure}

\begin{figure}[tb]
    \centering
    \includegraphics[width=\textwidth]{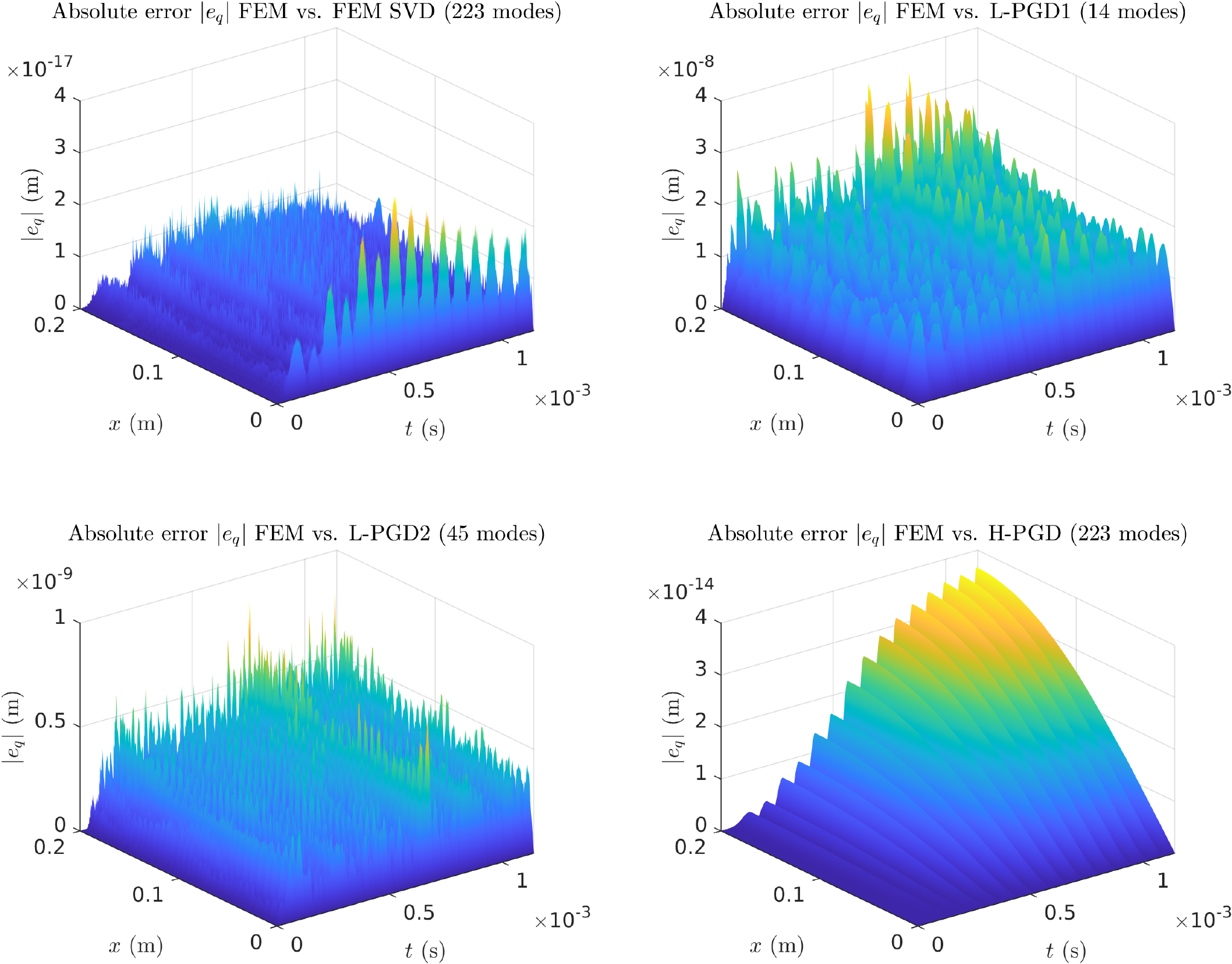}
    \caption{Case 5. Absolute errors in space and time between the reference displacement field and the SVD or PGD displacement field.} 
    \label{fig:damp-dir-neum-q}
\end{figure}

\begin{figure}[H]
    \centering
    \vspace{.2in}
    \includegraphics[width=\textwidth]{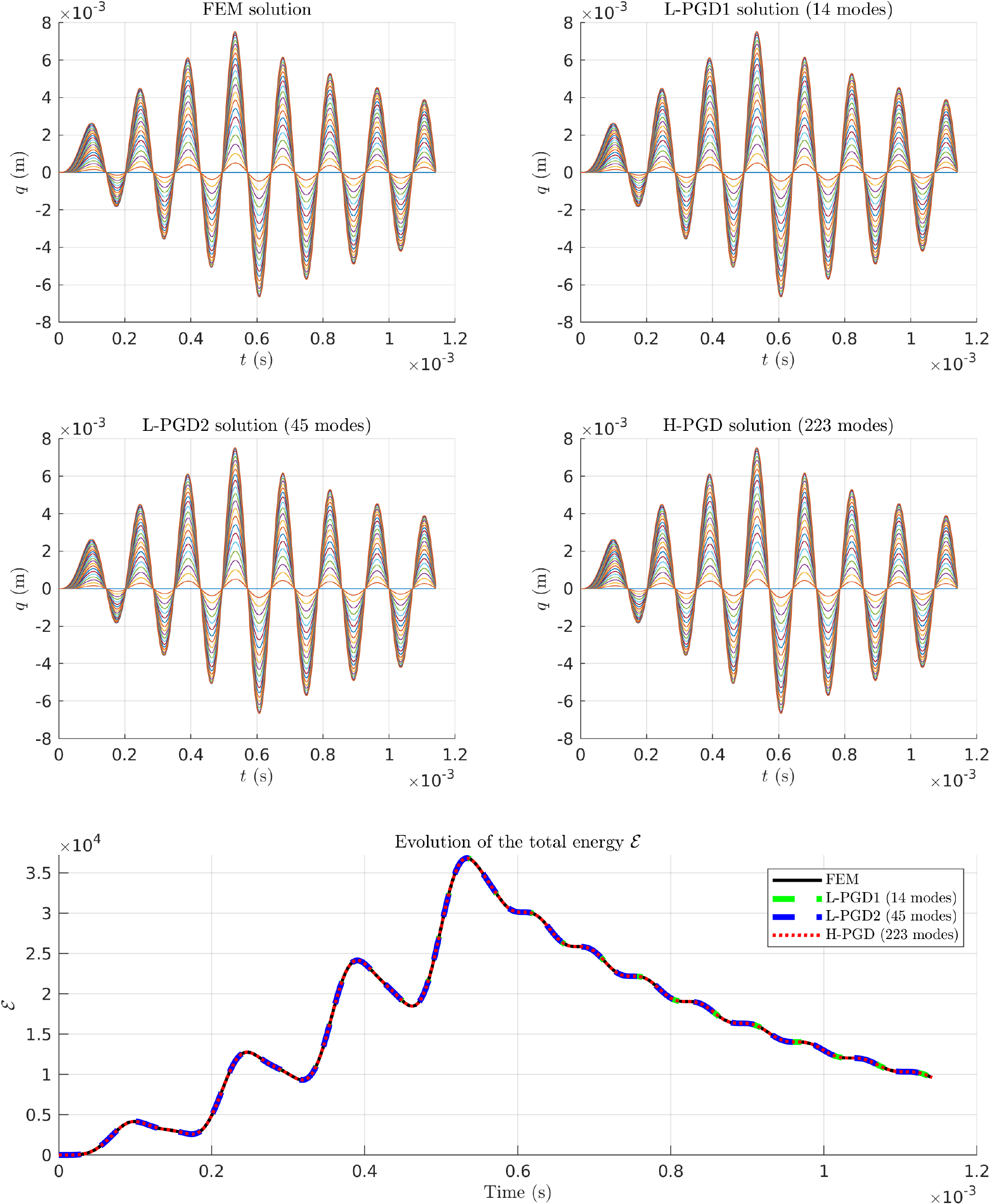}
    \caption{Case 5. (Top four plots) Evolution of the displacement field over time for the different reduction methods; the displacements are shown at 23 nodes uniformly distributed along the bar. (Bottom) Evolution of the total  energy of the reference and reduced systems versus time.} 
    \label{fig:damp-dir-neum-energy}
\end{figure}

The results are shown in Figures~\ref{fig:damp-dir-neum-plot},~\ref{fig:damp-dir-neum-q}, and~\ref{fig:damp-dir-neum-energy}. These are qualitatively very similar to those presented in Section~\ref{subsub:dir-neu}, except that the total energy of the bar decreases after $t\geq T/2$, as expected.
The main objective of this example is to illustrate that the H-PGD framework is also suitable for the study of linear elasticity problems accounting for energy dissipation. First, the H-PGD model reduction provides better stability and energy conservation of the original system than the L-PGD approaches. Moreover, reduced-order modeling methods for problems with damping based on modal decomposition~\cite{rixen} lead to an eigenvalue problem that requires a more elaborated treatment~\cite{zienkiewicz, ibrahimbegovic} than the eigenvalue problem obtained without damping. The Rayleigh hypothesis is often used to circumvent the issue introduced by the damping matrix~\cite{zienkiewicz}. However, the hypothesis does not have an unequivocal physical meaning~\cite{semblat, kareem} and may produce underdamped or overdamped behaviors in certain frequency ranges~\cite{zienkiewicz} (although this may be convenient in some cases). In contrast to the modal decomposition, it is not necessary in the PGD framework to resort to a special treatment in order to account for the damping term in the wave equation.

\subsection{Further discussion}


Non-symmetry and ill-conditioning of the matrices are issues that are also mentioned in~\cite[Pages 4 and 15]{boucinha2014}. The authors briefly discuss the eventual ill-conditioning of the operator $\mathbf{A}$ (non-symmetric), which represents the discretization of the space-time bilinear form of the problem. This operator is constant and its conditioning results from the fineness of the space-time discretization and the mechanical properties of the problem ($E$, $\rho$, etc.). In our case, the matrices $K_{\ell x}$, $M_{\ell x}$, $\Mb_{\ell x}$, $K_{h x}$ and $M_{h x}$ (Gram matrices) not only depend on the discretization and mechanical properties but are also non-constant. Indeed, their size grows with the number of enrichment $m$. Their ill-conditioning results mainly from the fact that the basis vectors may become linearly dependent. In conclusion, the nature of the matrices we studied is not the same as that of $\mathbf{A}$ and the reasons of their bad conditioning are different.


From a numerical point of view, we would like to emphasize that the only numerical difference between the Lagrangian update (\ref{eq:midpoint_pgd_lag_update}) and the Hamiltonian one (\ref{eq:midpoint_pgd_ham_update}) is the scaling factor $\rho A$. One may wonder whether this factor has an influence on the convergence of the H-PGD solver and the conditioning of the system. Multiple tests were run in the case where all parameters of the problem were set to unity: $E = \rho = A = \ell = 1$.

In this case, the scaling factor is equal to unity and the systems~(\ref{eq:midpoint_pgd_lag_update}) and~(\ref{eq:midpoint_pgd_ham_update}) are numerically identical. However, H-PGD remains numerically stable and still provides more accurate solutions than L-PGD. Therefore, it is not just the Hamiltonian formulation as a standalone formalism that enables improvements over the Lagrangian approach, but also the new possibilities that it offers in terms of algorithmic design.

Moreover, it is worth noting that the numerical values of the Young's modulus~$E$ and the density~$\rho$ play a major role in the non-symmetry. The non-symmetry is due to the temporal derivation. In other words, non-symmetry dominates if inertial effects are preponderant, i.e.\ when $\rho > E$ or, similarly, when the velocity $c=\sqrt{{E}/{\rho}}$ is small. One can find the same reasoning with the Heat Equation and the thermal diffusivity in~\cite{chinesta} (Page 68). Yet, in structural dynamics, $E \gg \rho$ in general. For instance, in our test cases, the material properties are chosen as those of steel with $E=220$~GPa, $\rho = 7000$~kg/m\textsuperscript{3}, similarly to~\cite{boucinha2014}. These numerical values are actually advantageous because they are not favorable to non-symmetry. Thus, the case of unitary parameters also enabled us to test our algorithm in examples where matrices become asymmetric and validate its robustness.

Energy conservation is related to the time integrators. Nevertheless, the computed PGD modes also play a significant role. This is best illustrated in the case of the problem with a Neumann boundary condition (see Figures~\ref{fig:dir-neum-plot} and~\ref{fig:damp-dir-neum-plot}). In particular, we observe on the top right plot that after the 5\textsuperscript{th} enrichment, L-PGD fails at recovering the modes with respect to the momentum field (i.e., error $\epsilon_p$). This results in a bad energy conservation. We conclude from these results that energy conservation is also sensitive to the PGD formulation and that H-PGD performs better that L-PGD by several orders of magnitude (as shown on the bottom left plot of the figures).

\section{Conclusion}
\label{sect:conclusions}

Galerkin-based PGD formulations based on the Hamilton's weak principle have been developed to derive reduced-order models of second-order hyperbolic systems. One of the objectives was in particular to compute a reduced model that preserves the energy of the system by means of stable, energy conservative integration schemes.
We have considered in this work two approaches, namely the L-PDG and the H-PGD. The former is based on the Lagrangian formalism while the latter follows from the Hamiltonian formalism. The H-PGD approach describes the system dynamics in terms of the generalized coordinates and the generalized momenta. The two fields can then be represented as two distinct expansions whose modes are solutions of coupled problems. The procedure brings some flexibility, in particular, it enables one to design orthogonalization and updating processes that ensure computational stability. Moreover, we have designed an adaptive fixed-point algorithm for the H-PGD approach that controls the convergence of the two fields separately. The combination of these procedures does improve convergence and eliminates redundancy in the enrichment modes of the H-PDG approach. As a result, for the test cases considered here, the H-PGD formulation showed a much better behavior in terms of stability and energy preservation than the L-PGD formulation. Finally, the Hamiltonian framework would allow one to recast the problem as a minimization problem with respect to the total energy of the system, i.e.\ the Hamiltonian, which makes it a suitable framework for the adaptive construction of goal-oriented PGD models as described in~\cite{kergrene}. This will be the subject of a future work.

\vfill
\section*{Declaration of competing interest}
\noindent
The authors declare that they have no known competing financial interests or personal relationships that could have appeared to influence the work reported in this paper.

\section*{Acknowledgements}
\noindent
Serge Prudhomme is grateful for the support by a Discovery Grant from the Natural Sciences and Engineering Research Council of Canada [grant number RGPIN-2019-7154].

\newpage
\bibliographystyle{unsrt}
\bibliography{literature.bib}

\end{document}